\pdfoutput=1
\documentclass[11pt,twoside,a4paper,cmspaper,final,collab]{cms-tdr}
\def\svnVersion{f8ae897}\def\svnDate{2023/08/04}\def\cmsCernNoTag{CMS NOTE-2023/009}\def\cmsCernDate{\today}\def\cmsMessage{Published in Computing and Software for Big Science as \href{https://doi.org/10.1007/s41781-024-00118-z}{doi:10.1007.s41781-024-00118-z} }
\begin{document}

\cmsNoteHeader{ECAL-ML4DQM}
\title{Autoencoder-based Anomaly Detection System for Online Data Quality Monitoring of the CMS Electromagnetic Calorimeter}
\date{\today}

\abstract{The CMS detector is a general-purpose apparatus that detects high-energy collisions produced at the LHC. 
 Online Data Quality Monitoring of the CMS electromagnetic calorimeter is a vital operational tool that allows detector experts to quickly identify, localize, and diagnose a broad range of detector issues that could affect the quality of physics data. 
 A real-time autoencoder-based anomaly detection system using semi-supervised machine learning is presented enabling the detection of anomalies in the CMS electromagnetic calorimeter data.
 A novel method is introduced which maximizes the anomaly detection performance by exploiting the time-dependent evolution of anomalies as well as spatial variations in the detector response. The autoencoder-based system is able to efficiently detect anomalies, while maintaining a very low false discovery rate. The performance of the system is validated with anomalies found in 2018 and 2022 LHC collision data. Additionally, the first results from deploying the autoencoder-based system in the CMS online Data Quality Monitoring workflow during the beginning of Run\,3 of the LHC are presented, showing its ability to detect issues missed by the existing system.}

\hypersetup{%
pdfauthor={Abhirami Harilal, Kyungmin Park, Manfred Paulini},%
pdftitle={Autoencoder-based Anomaly Detection System for Online Data Quality Monitoring of the CMS Electromagnetic Calorimeter},%
pdfsubject={CMS},%
pdfkeywords={CMS, ECAL, ML}}

\maketitle

\tableofcontents
\newpage

\section{Introduction}
\label{S:1}
\begin{figure}[bt]
\centering{
\includegraphics[width=0.8\textwidth]{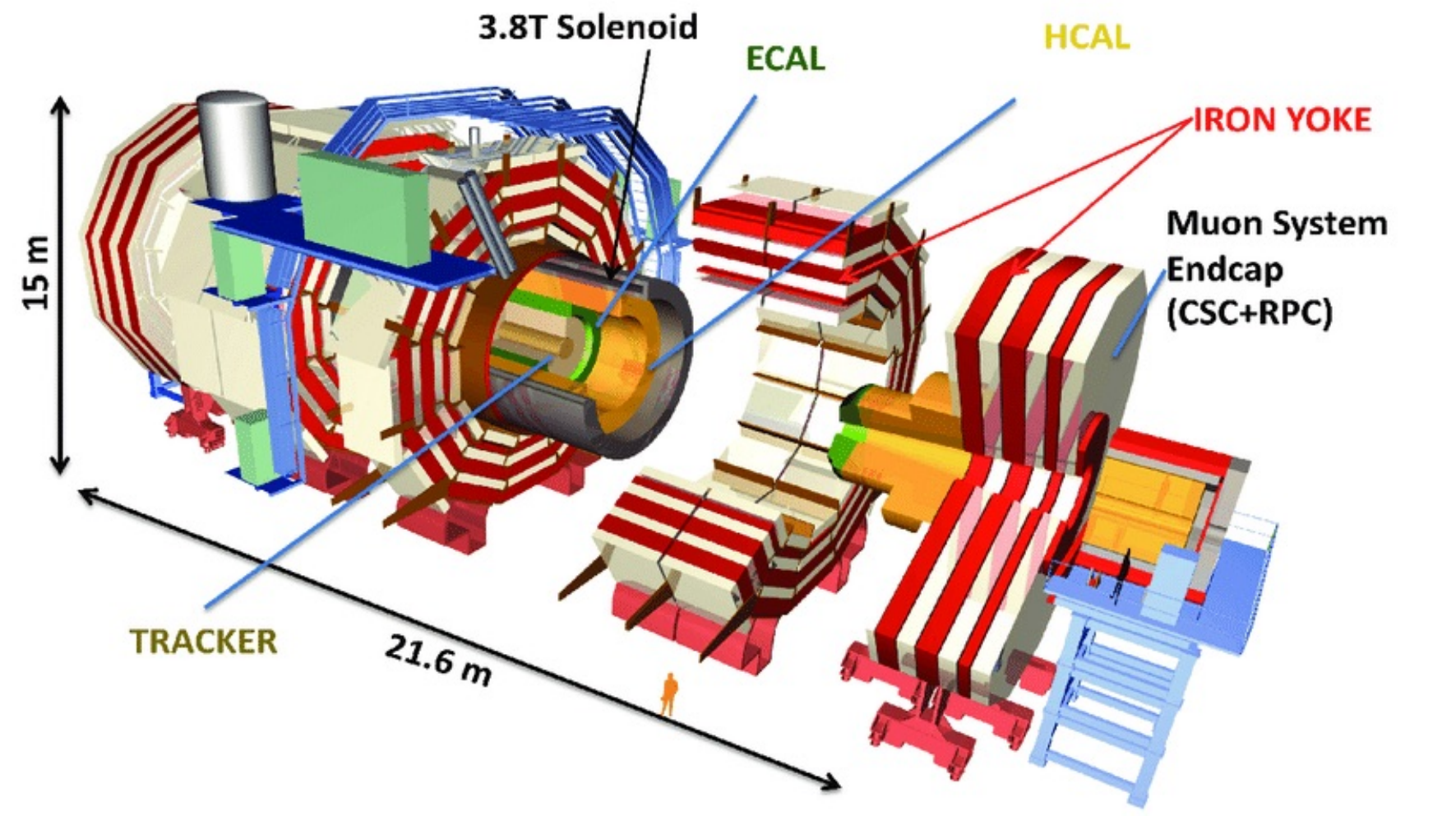}
}
\caption{Schematic view of the CMS detector and its various subdetectors.}
\label{fig:CMS}
\end{figure}

The CMS experiment has been taking high-quality proton-proton (pp)~collision data produced by the LHC at CERN for over a decade. Figure~\ref{fig:CMS} shows a schematic view of the CMS detector. The central feature of the CMS apparatus is a superconducting solenoid of 6\,m internal diameter, providing a magnetic field of 3.8\,T. Within the solenoid volume are a silicon pixel and strip tracker, a lead tungstate crystal electromagnetic calorimeter (ECAL), and a brass and scintillator hadron calorimeter, each composed of a barrel and two endcap sections. Forward calorimeters extend the pseudorapidity coverage provided by the barrel and endcap detectors. Muons are measured in gas-ionization detectors embedded in the steel flux-return yoke outside the solenoid. A more detailed description of the CMS detector, together with a definition of the coordinate system used and the relevant kinematic variables, can be found in Ref.~\cite{CMS:2008xjf}.

The energy of photons and electrons reconstructed in CMS events is obtained from the ECAL, the detector element most relevant for this paper.
In the barrel section of the CMS ECAL, an energy resolution of about 1\% is achieved for unconverted or late-converting photons in the tens of GeV energy range~\cite{CMS:Sirunyan_2021}.
An excellent ECAL detector resolution is the basis for precision physics measurements and the detection of new particles, as was the case for the discovery of the Higgs boson~H~\cite{ATLAS:2012yve,higgs_discovery}. For example, the diphoton mass resolution measured in the Higgs boson decays, H$\to\gamma\gamma$, is typically in the 1--2\% range, depending on the measurement of the photon energies in the ECAL and the topology of the photons in the event~\cite{CMS:2020xrn}. 
Precision measurements of particle decays such as H$\to\gamma\gamma$ do not only rely on  excellent detector calibrations but start with the recording of high-quality raw data. 

The CMS data quality monitoring (DQM) system~\cite{DQM2019_Azzolini} is a crucial operational tool to record high-quality physics data. Presently, the DQM consists of a software system that produces a set of histograms that are based on a preliminary analysis of a subset of data collected by the CMS detector. Conventional cut-based thresholds are used to define quality flags on these histograms which are  monitored continuously by a DQM shifter in the CMS control room who reports on any apparent irregularities observed.
While this system has proven to be dependable, the changing running conditions and increasing collision rates, together with aging electronics, bring forth failure modes that are newer and harder to predict. Machine learning (ML) techniques are nowadays widely used in high-energy physics~\cite{ML_whitepaper} and provide an excellent tool for anomaly detection in particle physics searches (see e.g. Ref.~\cite{Nachman-anomalydetection}).  Thus, ML is a natural choice for an automated system monitoring the data quality of an experiment. 
Previous efforts in ML for DQM within the CMS collaboration have explored such techniques~\cite{DT_AE_2019, ECAL_AE_2019}. 
In this paper, a semi-supervised method of anomaly detection for the online DQM of the CMS ECAL is presented, exploiting an autoencoder (AE)~\cite{AE} on ECAL data processed as two-dimensional~(2D) images.
In a novel approach, correction strategies are implemented to account for spatial variations in the ECAL response as well as the time-dependent nature of anomalies in the detector. 

While preliminary studies performed for the ECAL barrel~\cite{ECAL_AE_2019} showed the potential of detecting anomalies using AE for DQM, significant improvements in AE reconstruction and resolution are presented in this work together with novel post-processing strategies based on physics insights that enable the detection and localization of various anomalies with a very low false detection rate. 
The ML-based system is deployed in the online DQM for the ECAL during LHC Run\,3 collisions, complementing the existing DQM plots. 
The first results indicate that the AE-based anomaly detection system is a highly valuable diagnostic tool for ECAL experts involved in real-time data taking operations.

This paper is organized as follows: Section~\ref{sec:experimentalEnv} introduces the CMS ECAL and its current DQM system, and Section~\ref{sec:strategy} presents the AE network and the AE-based anomaly detection strategy, including the datasets used and their pre-processing. 
Section~\ref{sec:ML_part} discusses the strategy of training and validation of the network, as well as the corrections that are applied to account for the deviations of the ECAL response as a function of position and time. A metric to assess the performance of the AE-based anomaly detection method is also described, and a comparison to a baseline scenario similar to the existing ECAL DQM system is explained. Section~\ref{sec:Results} presents the AE performance on validation and test datasets with anomalies. Section~\ref{sec:deploy} discusses the deployment of the real-time AE-based anomaly detection system in the Run\,3 online ECAL DQM operation, followed by a summary in Section~\ref{sec:summary}.
 
\section{Experimental Environment}
\label{sec:experimentalEnv}
\subsection{Proton-Proton Collisions at the LHC}

The LHC has provided pp~collisions at center-of-mass energies of 7 and 8\,TeV (2009--2012), rising to 13\,TeV (2015--2018), and further to 13.6\,TeV for the ongoing Run\,3 that started in 2022. 
Each LHC beam consists of about 2500 tightly packed bunches of $\sim\!10^{11}$~protons with 25\,ns bunch spacing. The beams travel in opposite directions in the beam pipe and collide at four intersection points, one of them being the center of the CMS detector.
LHC operations involve ``fills'', where a fill is defined as a period during which the same proton beams are circulating in the LHC, and it typically consists of ten or more hours of collisions. 
The collision rate expressed through the instantaneous luminosity varies during each fill, and it decreases with time as the number of particles in the proton bunches and the beam intensity decay. 
Additional pp~interactions within the same bunch crossing, referred to as ``pileup'' (PU), can contribute to additional tracks and calorimetric energy depositions, increasing the event activity in the detector. The PU is correlated with the instantaneous luminosity and is  thus higher at the beginning of a fill than at the end.

An LHC fill is often divided into CMS ``runs'' that correspond to a start and stop of the CMS data acquisition system. Each run is further divided into time intervals called ``lumi-sections'' (LS) of an approximate time duration of 23 seconds corresponding to $2^{18}$ LHC orbits, over which the instantaneous luminosity is considered to remain approximately constant.

\subsection{The CMS Electromagnetic Calorimeter}
\label{sec:CMSECAL}

\begin{figure}[bt]
\centering{
\includegraphics[width=0.8\textwidth]{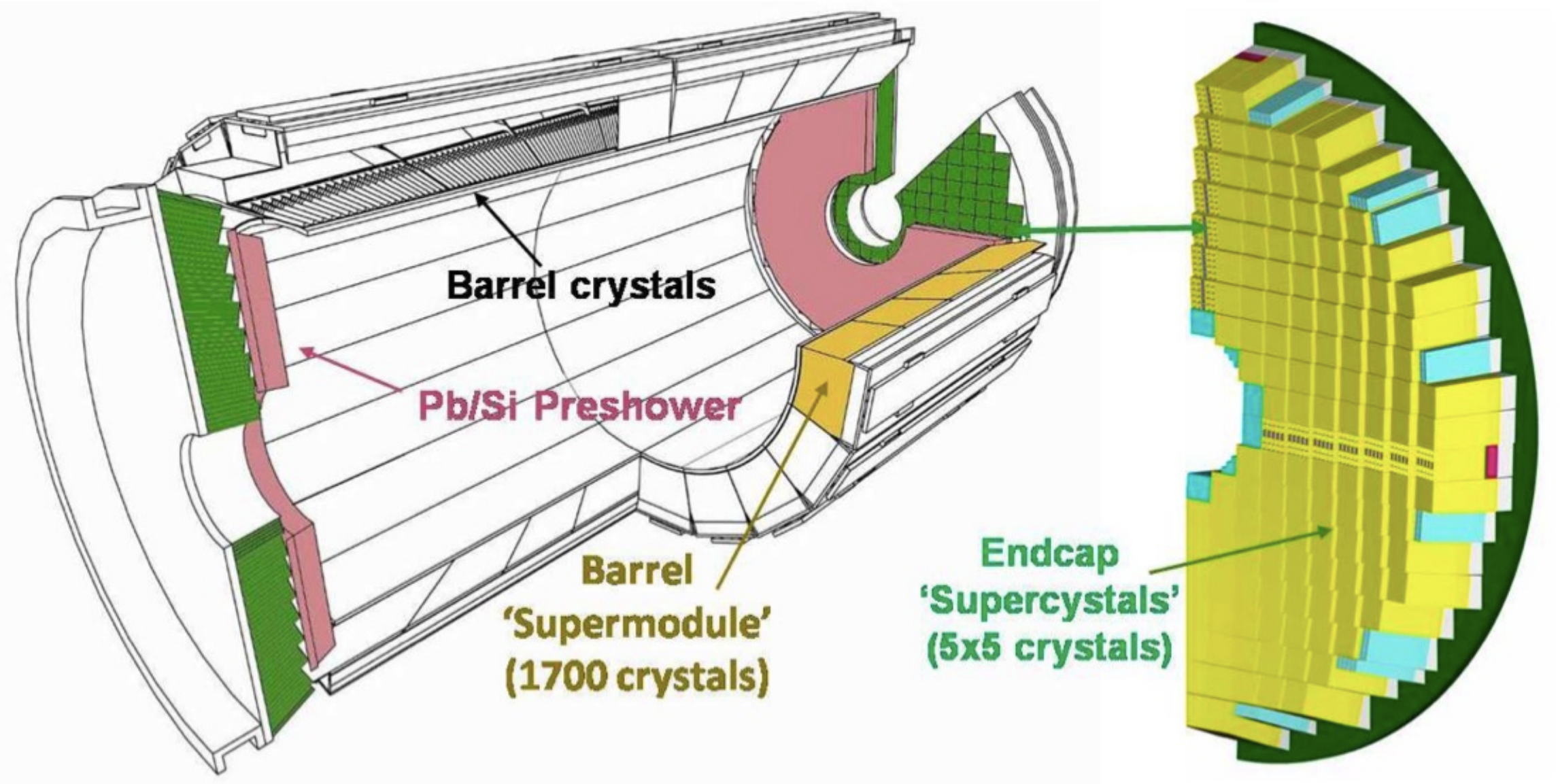}
}
\caption{Schematic view of the ECAL showing the cylindrical barrel closed by the two endcap regions with one half endcap displayed.}
\label{fig:ECAL}
\end{figure}

The CMS electromagnetic calorimeter provides homogeneous coverage in pseudorapidity $|{\eta}|<1.48 $ in a barrel region (EB) and $1.48 < |{\eta}| < 3.0$ in two endcap regions (EE$+$ and EE$-$) as shown in Fig.~\ref{fig:ECAL}. Preshower detectors consisting of two planes of silicon sensors interleaved with three radiation lengths of lead are located in front of each endcap detector. The ECAL consists of 75\,848 lead tungstate (PbWO$_4$) crystals. The barrel granularity is 360-fold in $\phi$ and (2×85)-fold in $\eta$ provided by a total of 61\,200 crystals, with each crystal having a dimension of 0.0174$\times$0.0174 in $\Delta \eta\times\Delta \phi$ space, while each endcap is divided into two halves, with each comprising 3662 crystals.

\begin{figure}[tbh]
\centering{
\subfloat[]{\includegraphics[width=0.6\textwidth]{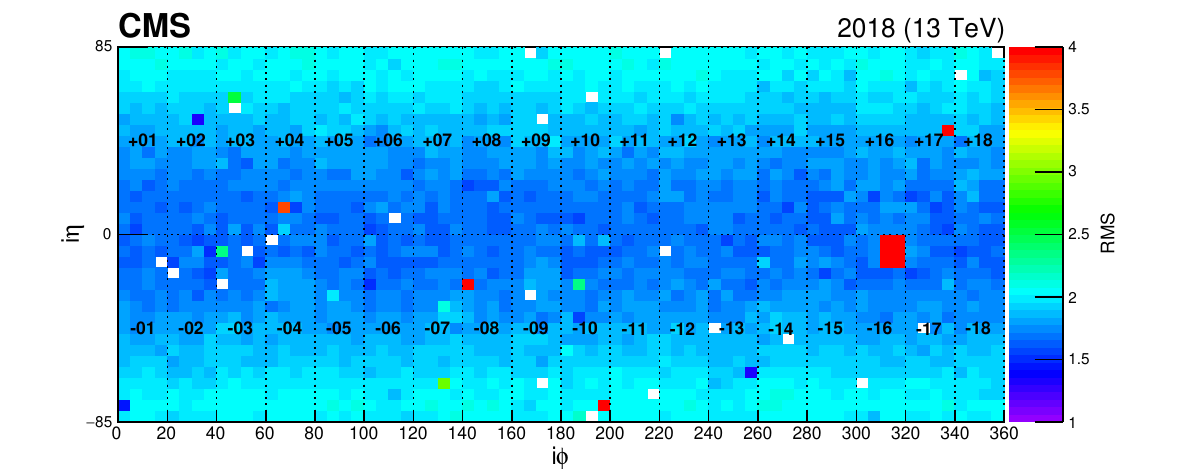}}
\subfloat[]{\includegraphics[width=0.3\textwidth]
{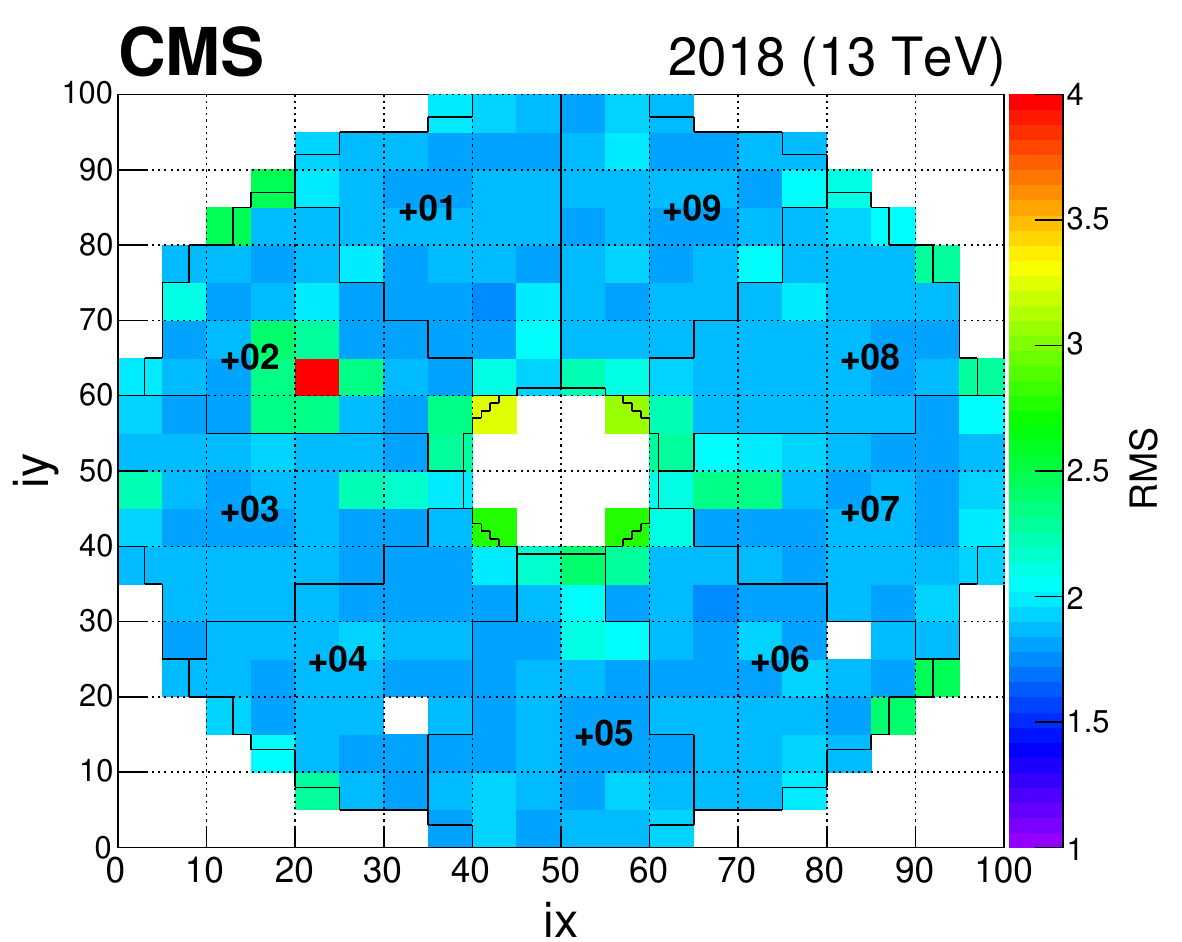}}
\\
\subfloat[]{\includegraphics[width=0.6\textwidth]{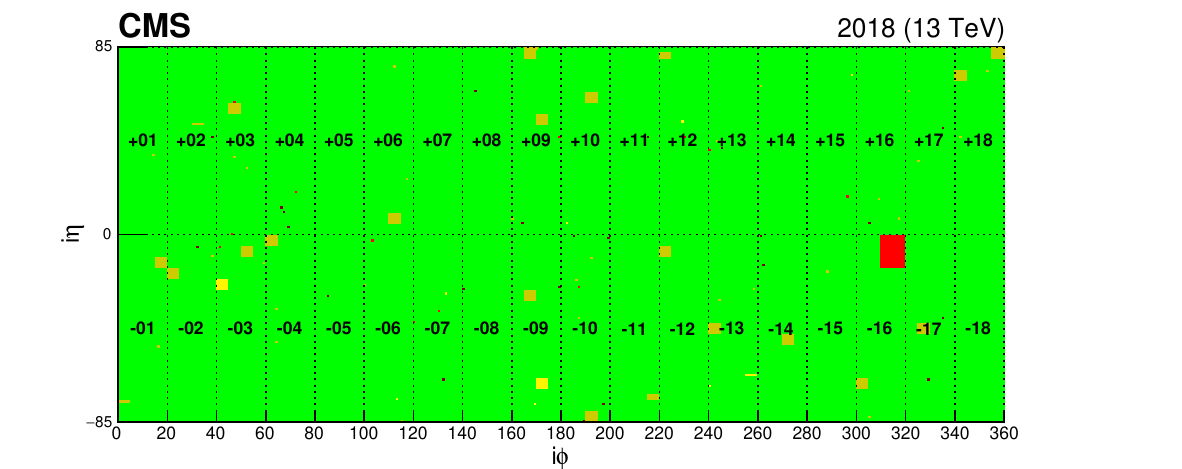}}
\subfloat[]{\includegraphics[width=0.3\textwidth]{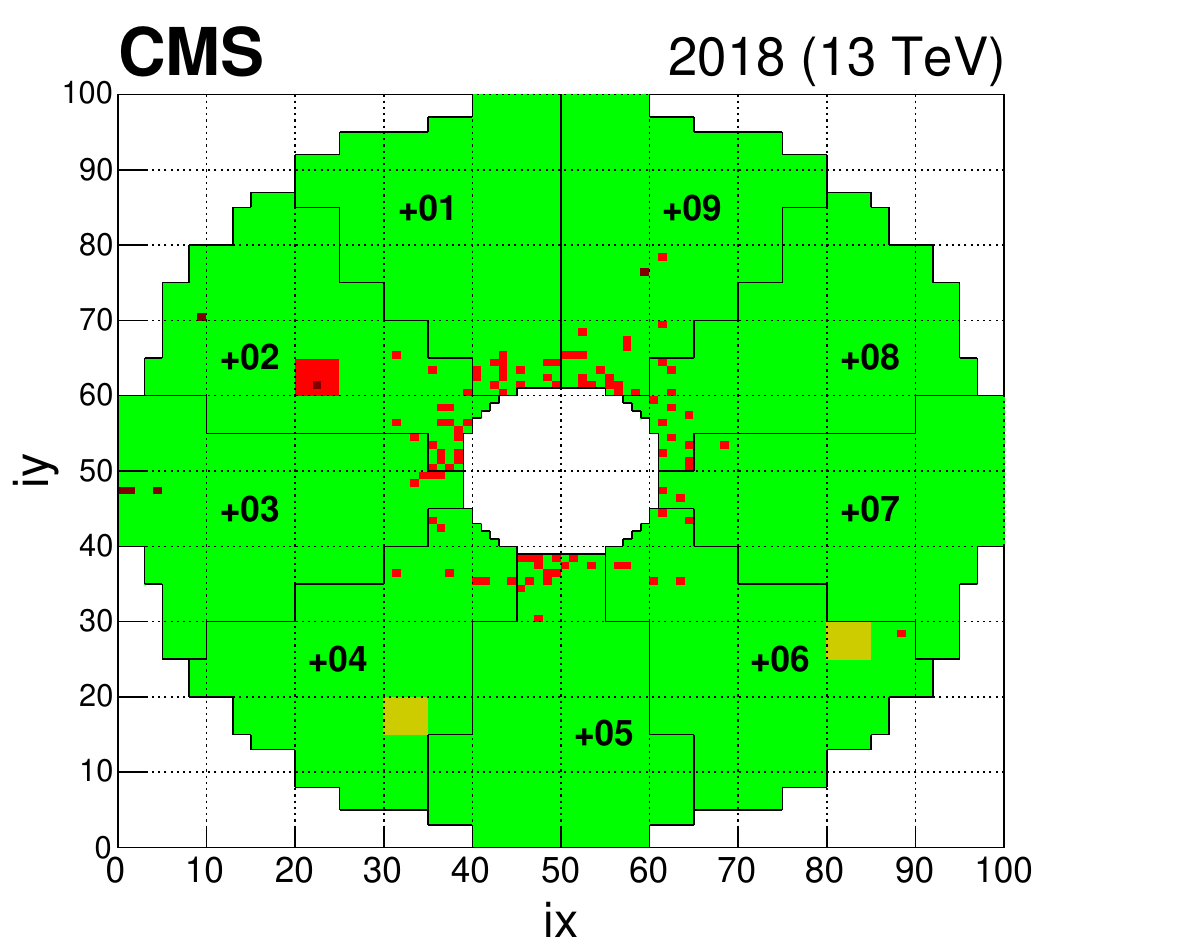}}
}
\caption{Example histograms from the ECAL DQM with (a) and (b) showing the distribution of RMS of the pedestal values in the barrel and EE$+$, respectively, drawn at a tower-level granularity. Diagrams (c) and (d) show the corresponding quality map for the two regions, drawn at a channel-level granularity, after a set of cuts is applied on the noise values shown in (a) and (b).}
\label{fig:task}
\end{figure}

Light signal from the ECAL crystals is detected, amplified, and digitized every 25\,ns. The collected data is stored in on-detector buffers awaiting an Accept/Reject signal from the first stage of the global CMS trigger system~\cite{CMS:2020cmk,CMS:2016ngn}. Upon receiving the Accept signal, a series of ten consecutive digitized samples are read out for each channel (or crystal) to measure the signal pulse amplitude and timing, after applying certain predetermined thresholds, to get an optimal pulse reconstruction as described in~\cite{CMS:2020reco}. 
These ten samples for each channel are referred to as digitized hits or ``digis''.
When a digi is registered for a given crystal, an occupancy value of 1 is counted for the crystal in any given DQM histogram.
A ``readout tower'' (henceforth referred to as a tower) is defined as a set of 5$\times$5 crystals (``supercrystals''), and 68 of these towers form a ``supermodule'' in the barrel (see Fig.~\ref{fig:ECAL}).
An unrolled projection of the ECAL barrel as well as one endcap is displayed in Fig.~\ref{fig:task}. 
Each single square in Figures~\ref{fig:task}(a)~and~\ref{fig:task}(b) represents a tower. The numbered rectangular regions in Fig.~\ref{fig:task}(a) represent the supermodules in the barrel, while the numbered regions in Fig.~\ref{fig:task}(b) indicate sectors in the endcaps.

The transverse energy of a trigger tower is computed by the front-end electronics and Trigger Concentrator Cards (TCC) within the off-detector electronics in the ECAL.  A classification flag is assigned by the TCC based on the threshold level that has been passed by the energy in each trigger tower.
A dedicated hardware device, called the Selective Read-out Processor~\cite{SRP}, receives from the TCCs the trigger tower energy maps and the corresponding flags, and it executes a selection algorithm~\cite{SRPalgo} that classifies the trigger tower as one of the following: ``suppressed'' (energy is below a low threshold), ``single'' (energy is in-between the low and high
thresholds), and ``central'' (energy is above the high threshold). For a ``central'' trigger tower, 3$\times$3 or 5$\times$5 regions around it are classified as ``neighbors''. 
The Selective Read-out Processor transmits the selective read-out flag for each tower to the Data Concentrator Cards, which read out the crystals as follows: for crystals that form ``central'', ``neighbor'', or ``single'' trigger towers, all energy
samples are kept, while crystals belonging to ``suppressed'' trigger towers are ignored or read out with a high zero suppression threshold.

\subsection{Data Quality Monitoring in ECAL}

During data taking, monitoring the data quality is a crucial, time-sensitive task to ensure optimal detector performance and the recording of high quality data suitable for physics analyses.
The CMS DQM~\cite{DQM2019_Azzolini} has two main modes of operation: offline and online. The offline DQM gives a retrospective view of data processed with the full statistics, passing various offline processing chains. It is mainly used in CMS for data quality certification~\cite{DQM_LTuura_2010}. The data are manually certified as good or not-good by comparing against several reference distributions, expert knowledge of running conditions, and known issues.

The online DQM offers a real-time snapshot of a subset of the raw data by populating a set of histograms after a preliminary analysis of the data, followed by a data quality interpretation step.
These histograms are updated every LS and are accumulated over a run. They are monitored continuously by a DQM shifter who reports on any apparent irregularities observed and informs detector experts to identify and act in real-time on any related issues with the detector. 
The current online DQM has many built-in alarms based on the set of histograms and indicates the presence of errors in a way that makes it easy to spot any ECAL issues at a glance.

\begin{figure}[tb]
\centering{
\subfloat[]{\includegraphics[width=0.5\textwidth]{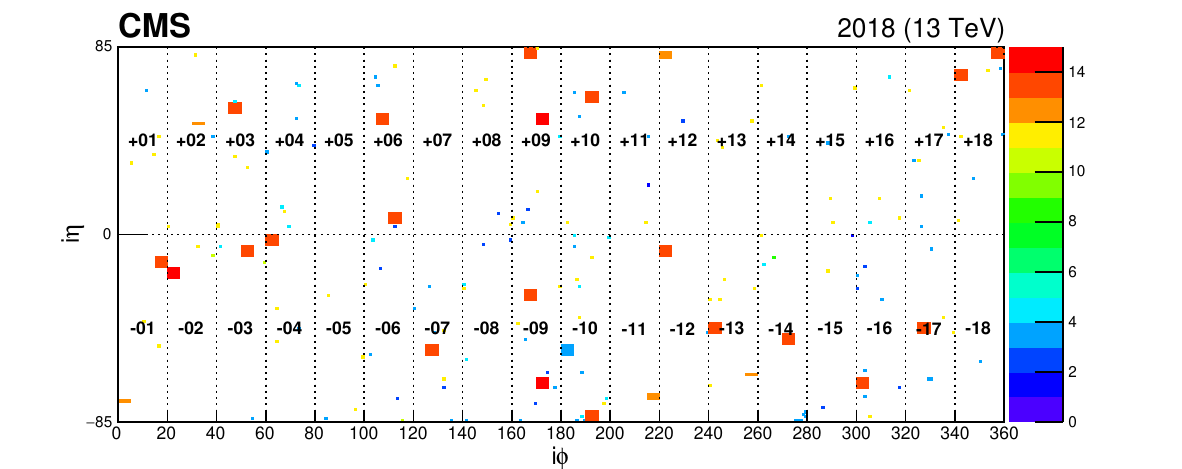}}
\subfloat[]{\includegraphics[width=0.25\textwidth]{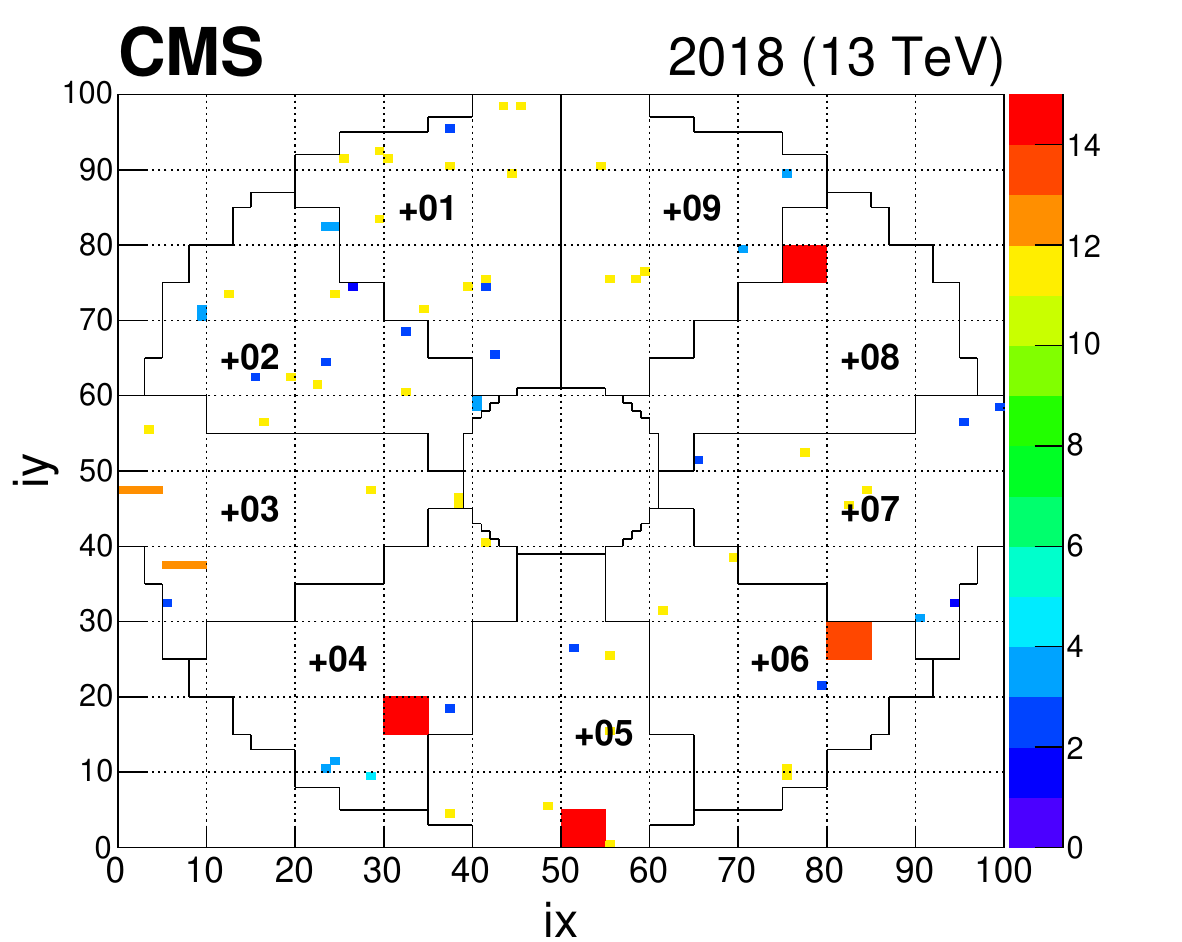}}
\subfloat[]{\includegraphics[width=0.25\textwidth]{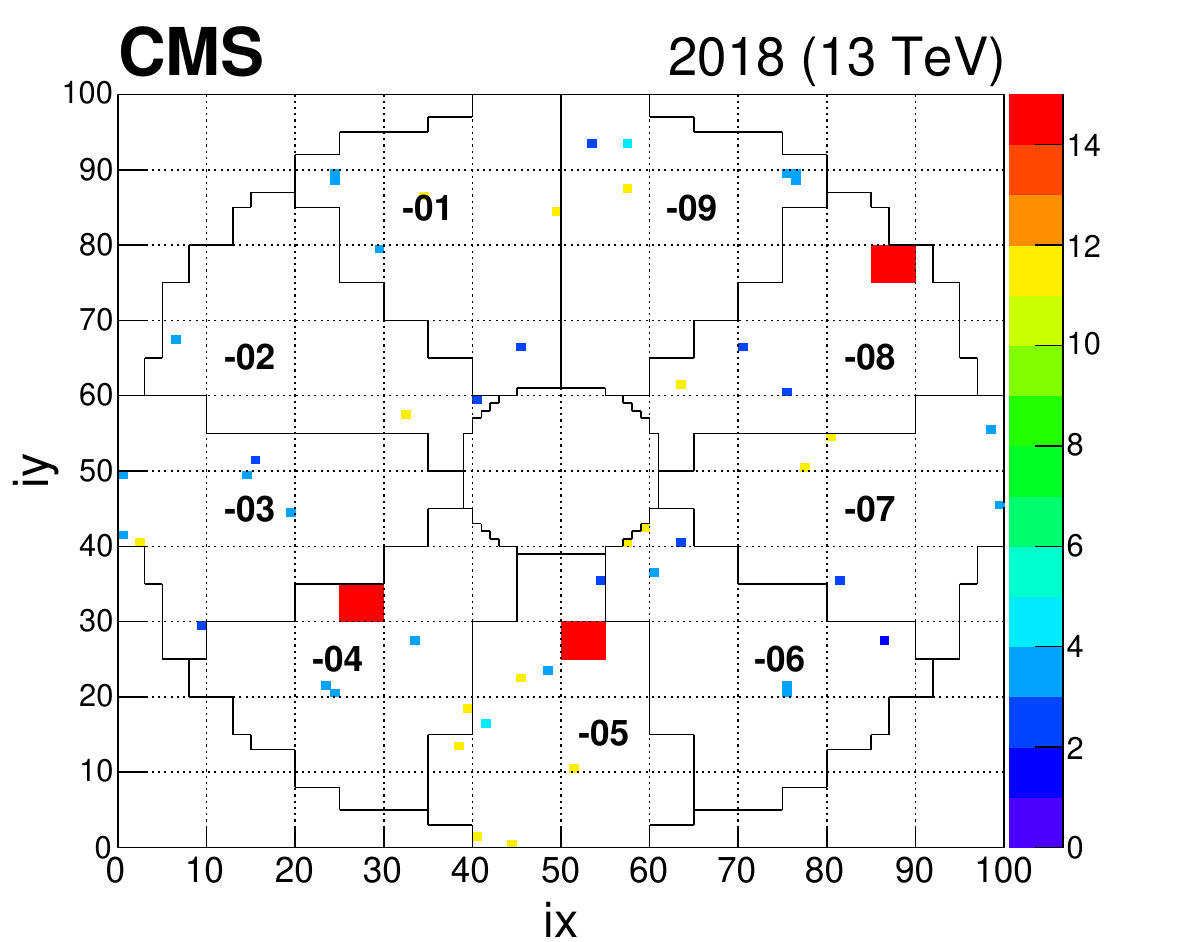}}
}
\caption{Channel status maps used in the ECAL DQM indicating the known problematic channels color coded for various types of errors for (a) EB, (b) EE$+$, and (c) EE$-$.}
\label{fig:channel_status}
\end{figure}

There are two kinds of histograms present in the ECAL DQM: ``Occupancy-style'' histograms shown in Figures~\ref{fig:task}(a)~and~\ref{fig:task}(b) filled with critical quantities from the real-time detector data and ``Quality-style'' histograms displayed in Figures~\ref{fig:task}(c)~and~\ref{fig:task}(d). Quality-style histograms are obtained by applying predefined thresholds and requirements to the Occupancy-style histograms, where the thresholds are derived from typical detector response. In the examples shown in Fig.~\ref{fig:task}, the Occupancy-style histograms are plotted at a tower-level granularity while the Quality-style histograms are plotted at a channel-level granularity.
The quality histograms are drawn in easily identifiable colored maps, and the color code scheme used is as follows: green for ``good'', red for ``bad'', brown for ``known problems'', and yellow for ``no data'' (that may or may not be problematic depending on the context). Information about known bad channels and towers is displayed in a channel status map with an example given in Fig.~\ref{fig:channel_status}. Here, the different colors correspond to the status values as: 0 -- channel OK, 1 -- channel with pedestal not in range, 2 -- channel with no laser, 3-7 -- various types of noisy channels, 8-9 -- channels in fixed gain, 10-14 -- various types of dead channels, and status $>$ 14 -- channels with issues in low voltage or high voltage.
Depending on the severity of the problem, these channels are either masked in the Data Acquisition system of the detector or in the 
offline reconstruction, and the data from them are ignored and may or may not show up in the DQM. This information is stored and regularly updated in a database and is used to mark the towers in the DQM quality plots in dark colors, e.g. dark brown.

As an example of existing DQM plots, Figures~\ref{fig:task}(a)~and~\ref{fig:task}(b) show the distribution of the pedestal Root Mean Square (RMS) in EB and EE$+$, respectively. Regions with high noise could indicate, for instance, a potential problem with the high voltage in the detector. Figures~\ref{fig:task}(c)~and~\ref{fig:task}(d) indicate the corresponding quality maps. Here, the crystals are shown in red if their RMS values are greater than the set thresholds.

\begin{figure}[tb]
\centering{
\subfloat[]{\includegraphics[width=0.6\textwidth]{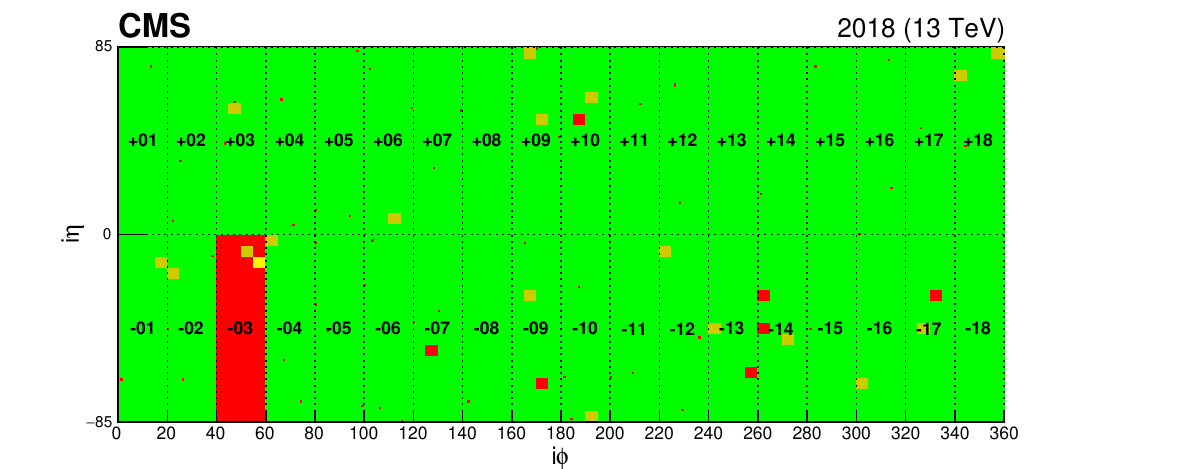}}
\subfloat[]{\includegraphics[width=0.3\textwidth]{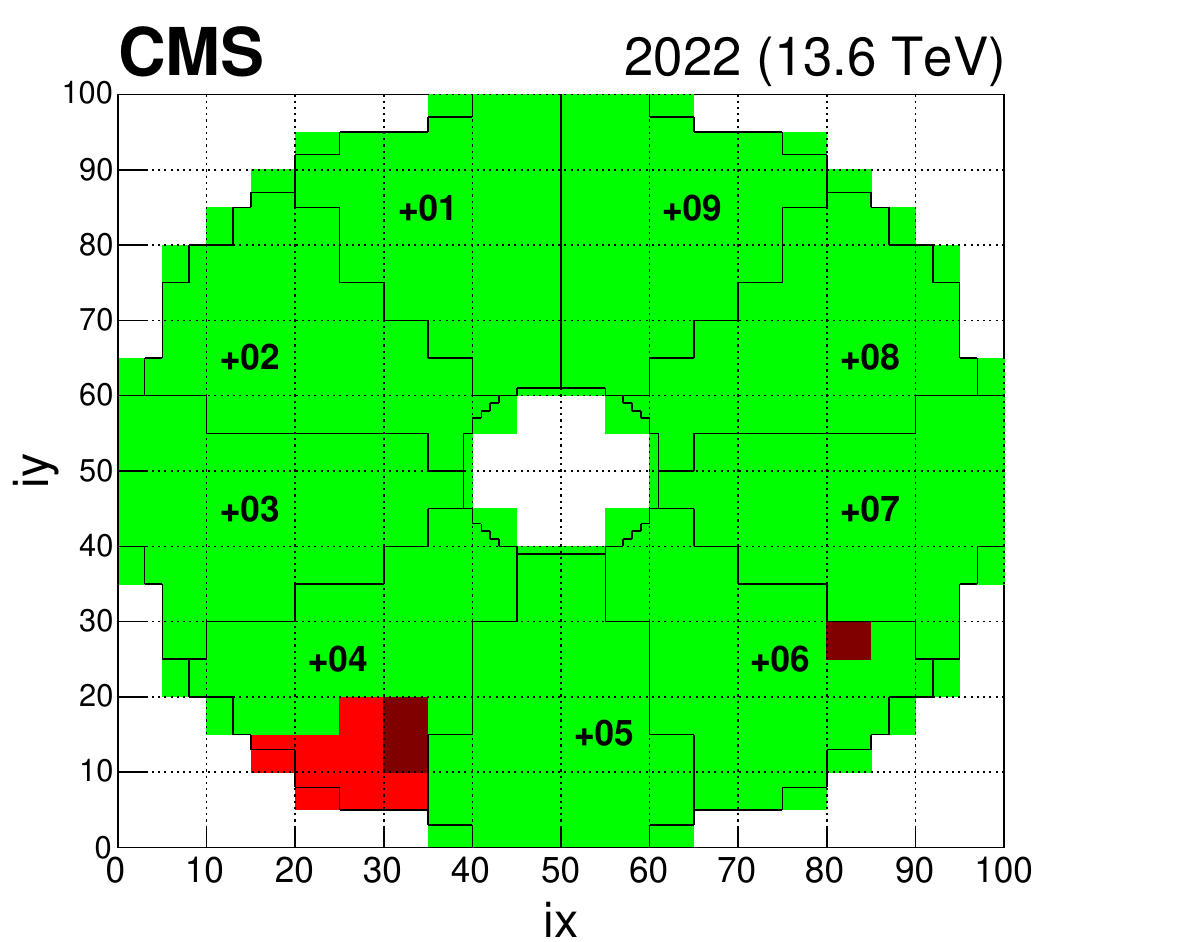}}
}
\caption{DQM quality plots with different anomalies shown in red, while the towers with known issues show up as dark brown or dark yellow. (a) EB$-$03 turned off due to a voltage failure and seen in red.
(b) Anomaly in EE$+$04  (marked in red) originating from an electronics failure affecting 200 channels. }
\label{fig:erro_ex}
\end{figure}

Anomalies can come in different shapes and sizes as illustrated in Fig.~\ref{fig:erro_ex}, attributed to various sources such as underlying hardware components.
Furthermore, the ever-changing LHC and CMS running conditions can often result in failure modes that are hard to predict.  
Although continuous improvements of the DQM have allowed the existing system to be updated and respond to new problems, e.g. issues with electronic components, it can become challenging to define hard-coded rules and thresholds manually for every failure mode. To overcome such challenges, an automated anomaly detection system using machine learning is developed to complement the existing DQM. 

\section{Machine Learning Based Anomaly Detection Strategy}
\label{sec:strategy}
Unsupervised or semi-supervised ML methods used for anomaly detection are an excellent choice to supplement the ECAL DQM system. With the existing manual data certification procedure in CMS and ECAL, the recorded offline data is certified to be good or bad for physics analyses and/or for detector calibration based on various defined markers. 
Using a semi-supervised approach, the network is trained exclusively on the certified good physics dataset, so that it learns the patterns of good data and is able to detect anything that differs from the nominal patterns it has learned.
The network is able to detect anomalies without the need to explicitly see the anomalous data during training.
A semi-supervised anomaly detection and localization method for the online ECAL DQM is developed using an AE network based on a computer vision technique. The AE is built with a convolutional neural network (CNN) architecture~\cite{lecun:98} exploiting ECAL data processed as 2D images. 
Corrections that take into account the spatial variations in the ECAL response and the time-dependent nature of anomalies in the detector are implemented in order to effectively maximize the anomaly detection efficiency while minimizing the false positive detection probability. 

\subsection{Autoencoder Network}
\label{sec:AE}
An AE is used with a Residual Neural Network (ResNet)~\cite{ResNet} CNN architecture implemented with a PyTorch~\cite{NEURIPS2019_9015} backend.
The encoder part of the AE takes the input data and compresses it into a lower dimensional representation, called the latent space, which contains a meaningful internal representation of the input data. The decoder part then decompresses the encoded
data back to the original image of the same dimensions, or reconstructs the image.
To measure how well the output matches the input, or the goodness of the AE reconstruction, a reconstruction loss~($\mathcal{L}$)~\footnote{Note that the reconstruction loss in this paper always refers to the AE reconstruction loss and has no relation to CMS particle reconstruction.} is computed using Mean Squared Error between the input ($x$) and the AE-reconstructed output ($x'$) as defined in Eq.~(\ref{eq:loss}).

\begin{equation}
\mathcal{L}(x,x') = ||(x-x')||^{2}
\label{eq:loss}
\end{equation}

A network trained on good images will learn to reconstruct them well by minimizing this loss function. When fed with anomalous data, the AE returns higher loss in the anomalous region, forming the basis of the anomaly detection strategy discussed in Section~\ref{sec:AnomalyStrategy}. 

\begin{figure}[bth]
\centering{
\includegraphics[width=0.7\textwidth]{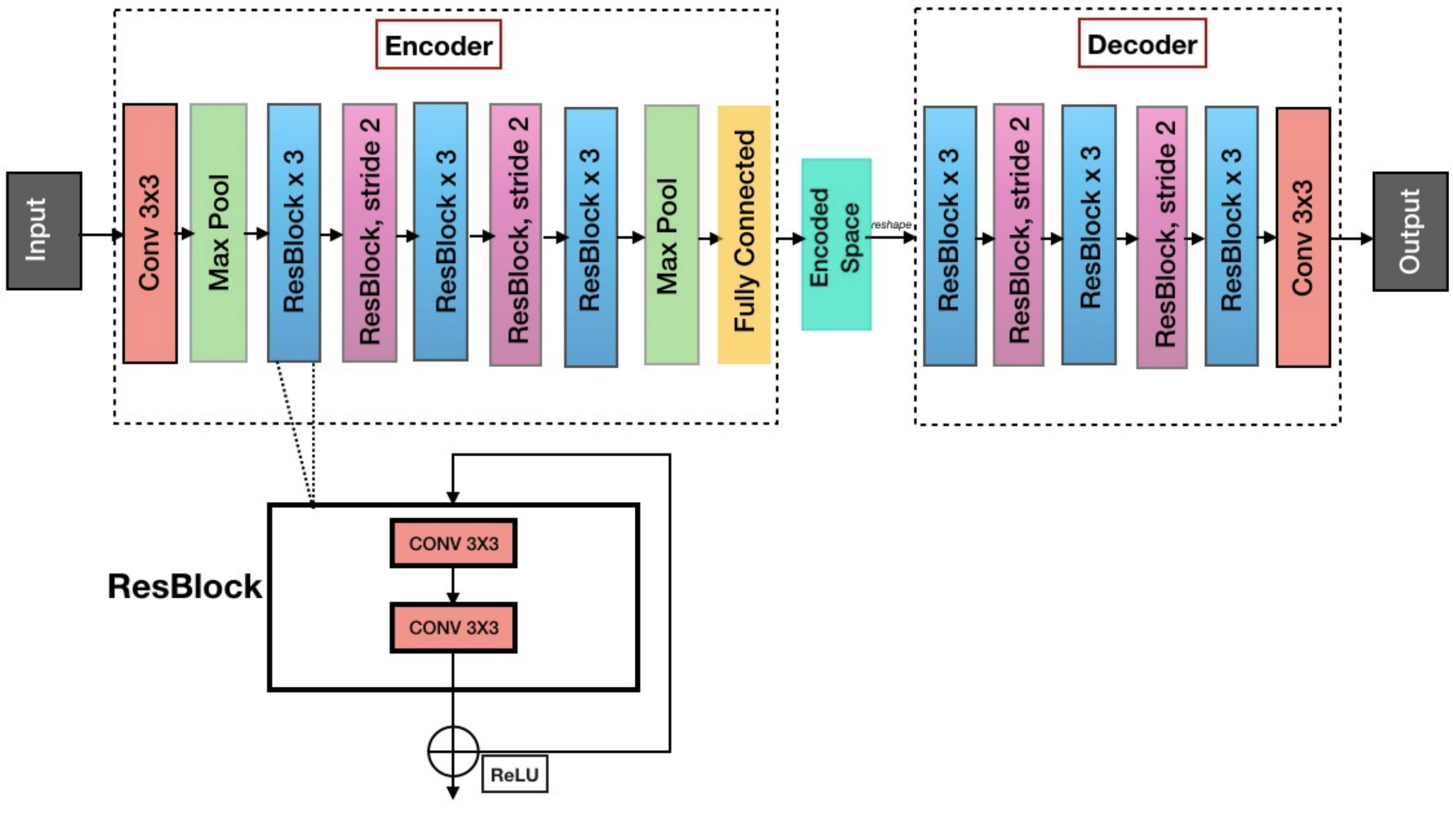} }
\caption{AE architecture showing the encoder and decoder networks, with the ResBlock structure displayed. }
\label{fig:AE-model}
\end{figure}

Figure~\ref{fig:AE-model} shows the architecture used for the AE.
Each ``ResBlock'' consists of two convolutional layers, with a Rectified Linear activation function (ReLU)~\cite{Nair:2010:RLU:3104322.3104425} in between and a residual mapping. The input image (shape of 36$\times$72 for EB and 22$\times$22 for EE) is passed through the encoder network that consists of a CNN, followed by a maxpool layer that aggregates the maximum values of the feature maps.
It is then sent through sequential layers of ResBlocks, where the feature maps are up-sampled progressively. This is followed by a global maxpool that creates a compressed dense layer of the encoded space. The encoded layer is then passed to the decoder network as the input, which reverses these operations and gives out a reconstructed image. Three separate models are trained with this architecture: one for the barrel and one for each of the two endcaps. The choice of training the separate models for the barrel and the endcaps is attributed to the differences in their shape, granularity, and response.

\subsection{Dataset and Event Selection}
\label{sec:dataset_eventSel}
The dataset used for training and validation of the AE network is taken from CMS runs collected in 2018 during LHC Run\,2. It contains runs that are manually certified as good by the data certification procedure in CMS and ECAL. Each input image for the AE is the digi occupancy map from a single LS that can be seen from Fig.~\ref{fig:occ_ex}.
It is to be noted that the current DQM checks a much larger phase space of detector quantities. Digi occupancy maps are chosen for the ML-based DQM as they are a good indicator of most detector problems.
The occupancy maps are processed offline to include 500 events per LS, which is approximately the number of data received per LS in the online DQM. Although the actual number of events varies per LS in real-time data taking, this approximation is used to ensure that the network is sensitive to variations in the occupancy due to anomalies and not due to differences in the collected statistics. It is demonstrated in Section~\ref{sec:deploy} that the AE-based system performs very well in the online DQM deployment on real-time data with varying numbers of events per LS.

\subsection{Detector Images}

\begin{figure}[tbh]
\centering{
\subfloat[]{\includegraphics[width=0.46\textwidth]{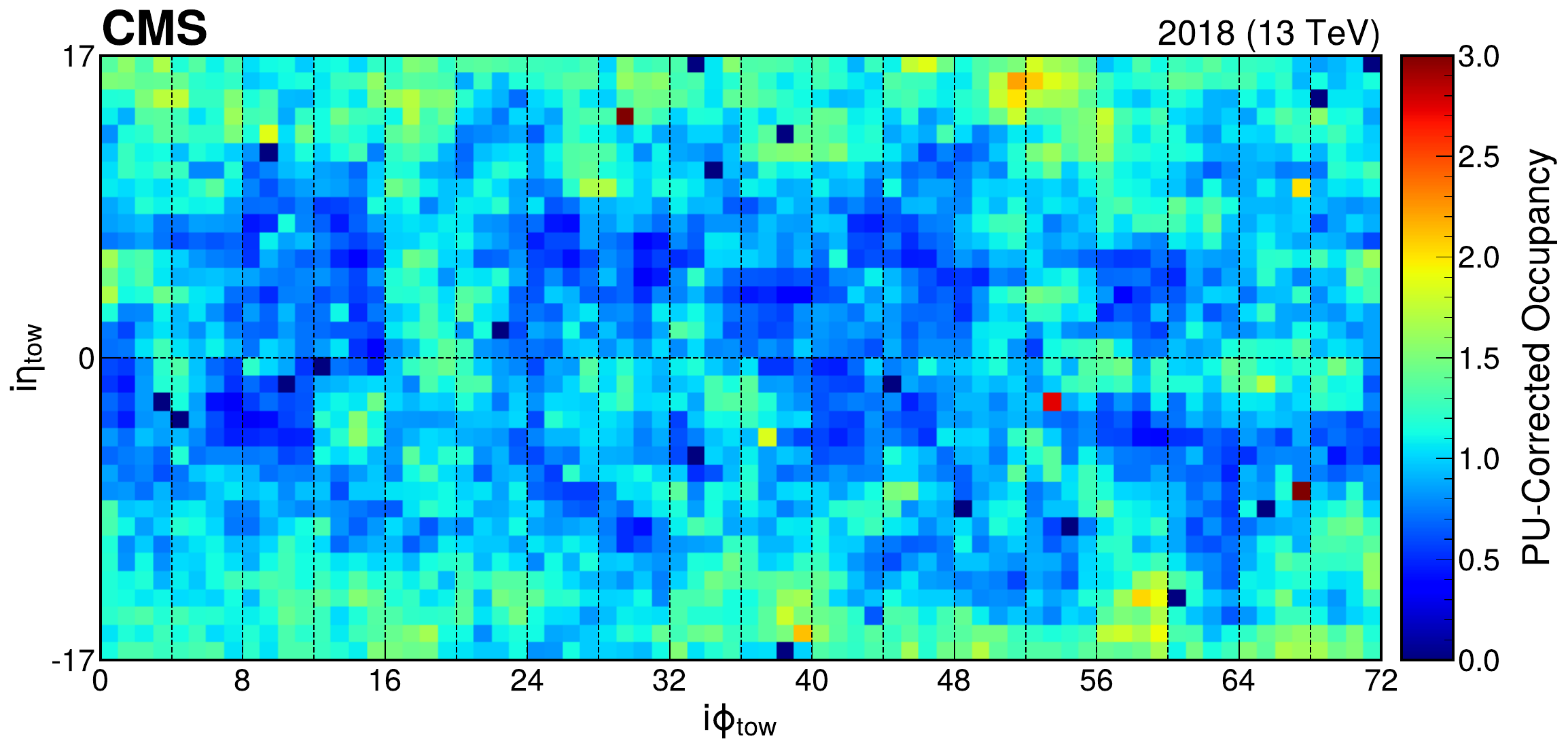}}
\subfloat[]{\includegraphics[width=0.25\textwidth]{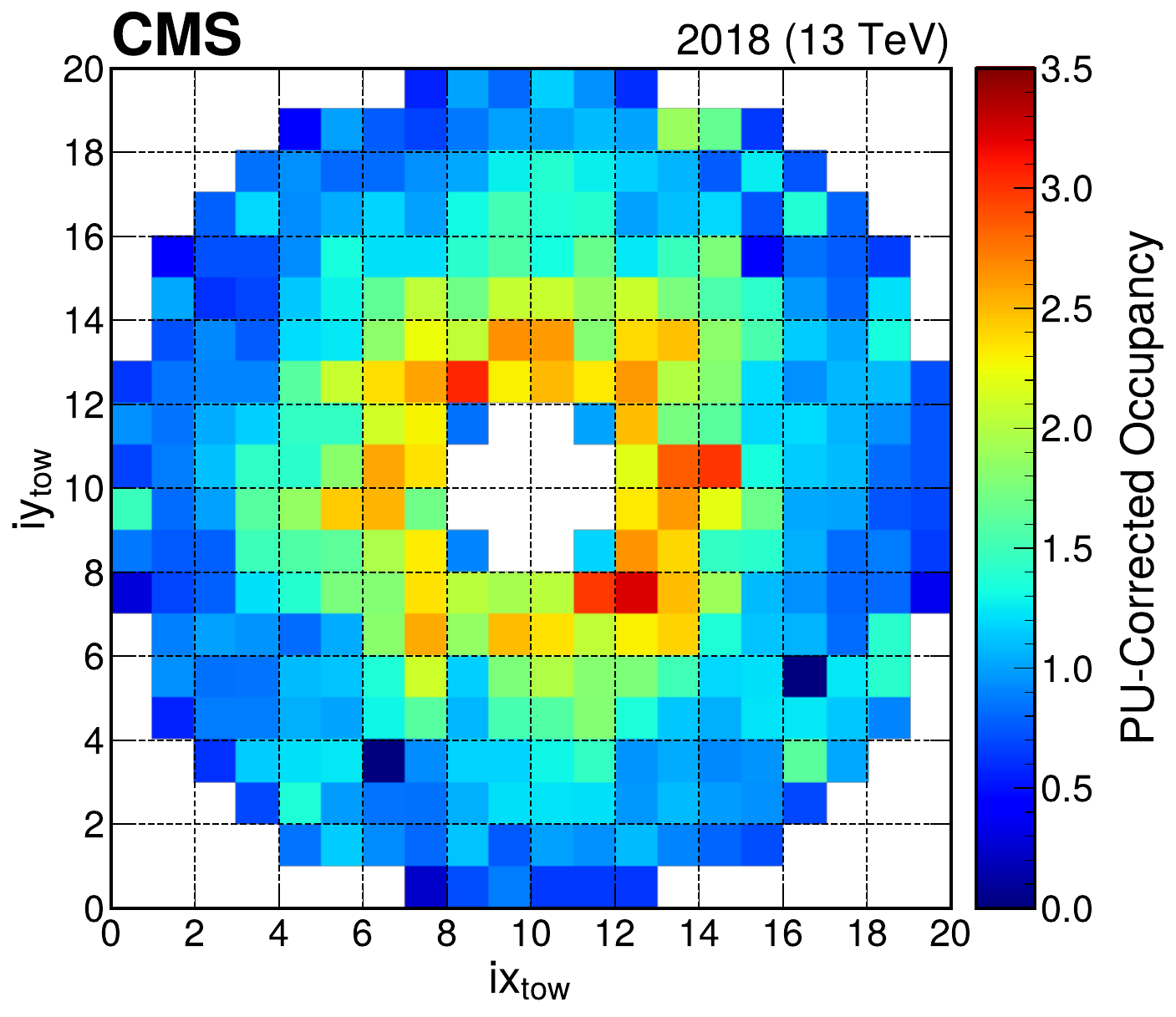}}
\subfloat[]{\includegraphics[width=0.25\textwidth]{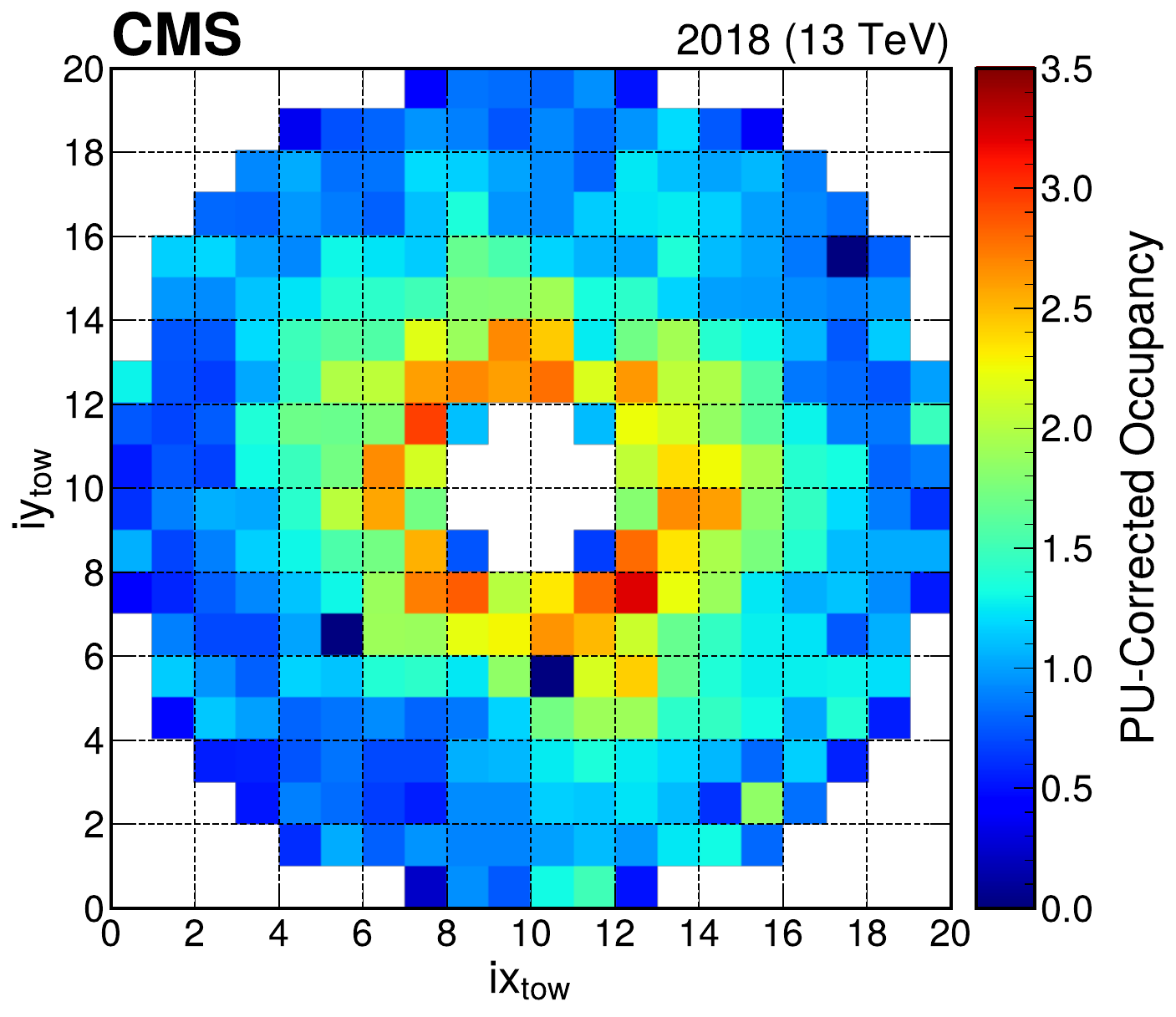}}
}
\caption{Typical occupancy maps for a single LS from the training dataset for (a) EB, (b) EE$+$, and (c)~EE$-$. During training, the edges of these maps are padded.}
\label{fig:occ_ex}
\end{figure}

Occupancy histograms using digis from the ECAL barrel and endcap sections are fed to the network as 2D images for each LS. 
The occupancy images are drawn at a tower-level granularity with the image shapes of 34$\times$72 for the barrel and 20$\times$20 for the endcaps. 
Typical occupancy images for a single LS are shown in Fig.~\ref{fig:occ_ex}. 
Different from the usual crystal-level indexing used in most ECAL DQM plots, the  occupancy maps displayed here have a modified tower-level indexing for convenience, $i\eta_{tow}$-$i\phi_{tow}$ for the barrel and $ix_{tow}$-$iy_{tow}$ for the endcaps.
For the empty regions in the endcap images, e.g. ($ix_{tow}$,$iy_{tow}$)=(0,0), their occupancy values are set to zero during training and the loss values of these regions are not taken into account during inference.

The input images to the network are padded at the edges to mitigate edge effects, since learning at the boundaries becomes sub-optimal due to under-representation of the edge values during convolution. After padding by duplicating the first and last rows of the image for the barrel, the input image shape is 36$\times$72. For the endcaps, padding both the first and last rows and columns gives a 22$\times$22 input image shape. During the inference, however, the original shape of the images without the padding is used. 

\subsection{Pre-processing}
\begin{figure}[tb]
\centering{
\subfloat[]{\includegraphics[width=0.4\textwidth]{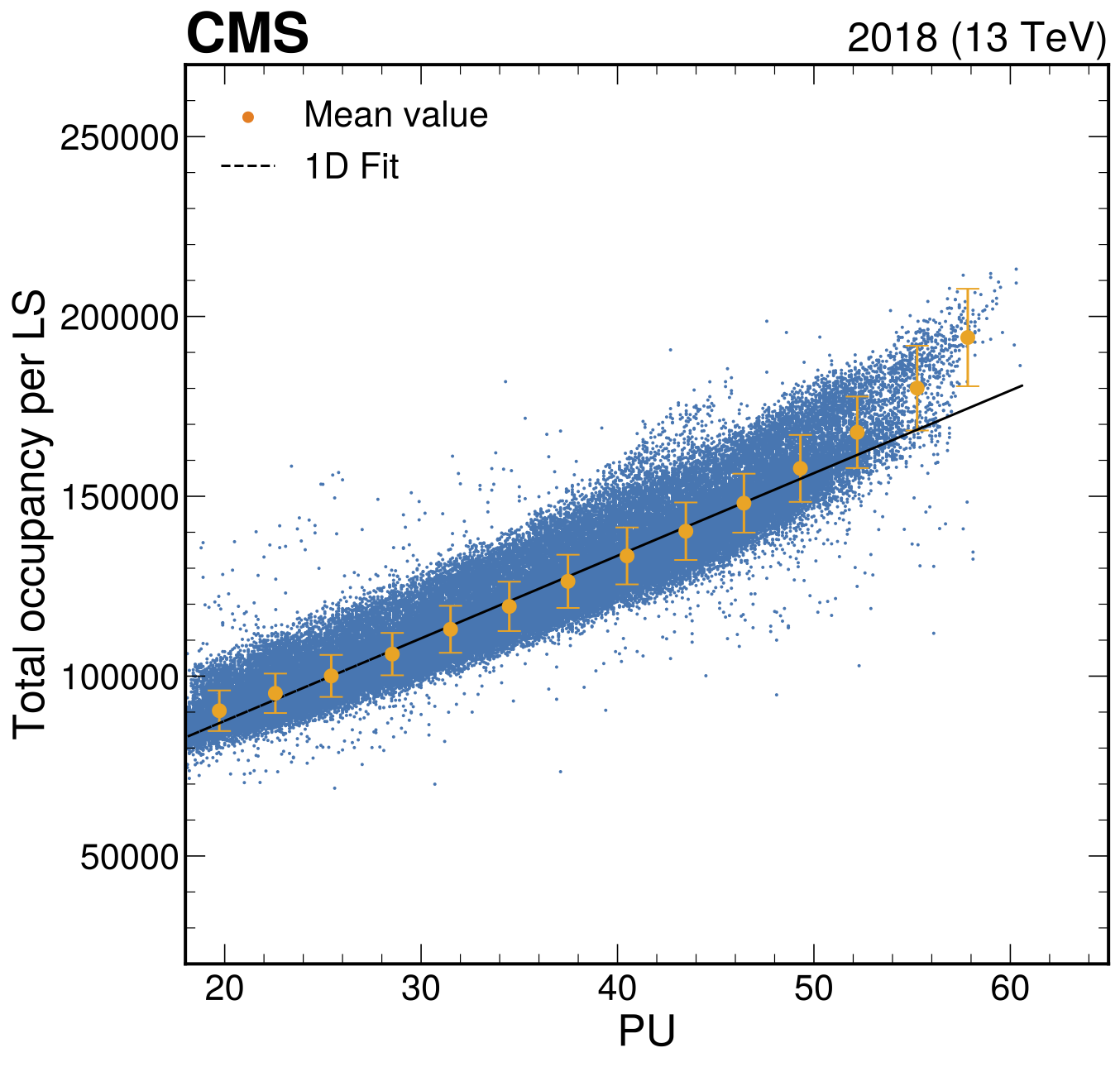}}
\hspace{1cm}
\subfloat[]{\includegraphics[width=0.38\textwidth]{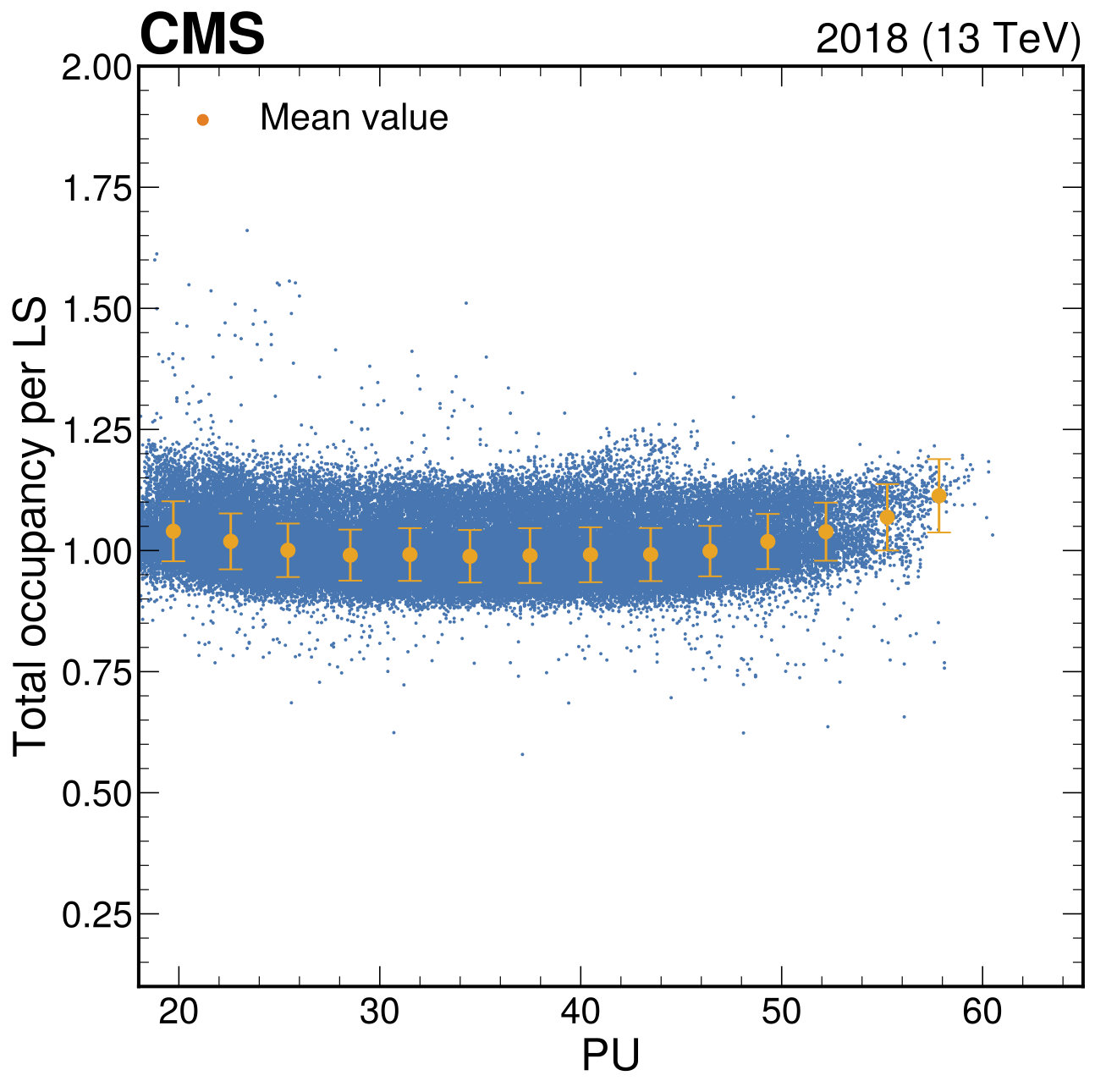}}
\caption{Sum of digi occupancy for each LS occupancy map versus PU (a) before and (b) after the PU correction is applied. Overlaid on the scatter plot are the mean values with standard deviation for each PU bin. }
\label{fig:fit}}
\end{figure}

To ensure consistent data quality interpretation across different LHC running conditions, it is important to find a coherent way of normalizing the occupancy to make it independent of factors such as the LHC instantaneous luminosity, which is correlated with PU.
As shown in Fig.~\ref{fig:fit}(a), the dataset used for training indicates a linear relation between occupancy and PU at first order.
A linear fit is performed to the distribution and a correction factor is derived from the fit parameters. After the correction is applied to each occupancy map per LS as pre-processing, an almost flat relation between occupancy and PU is obtained as seen in Fig.~\ref{fig:fit}(b). After removing the PU dependence across the dataset with the correction,
each occupancy map is then re-scaled such that the average occupancy across the towers is around one, giving typical occupancy maps as Fig.~\ref{fig:occ_ex}.

\subsection{Anomaly Detection Strategy}
\label{sec:AnomalyStrategy}
\begin{figure}[tb]
    \centering{
    \includegraphics[width=0.8\textwidth]{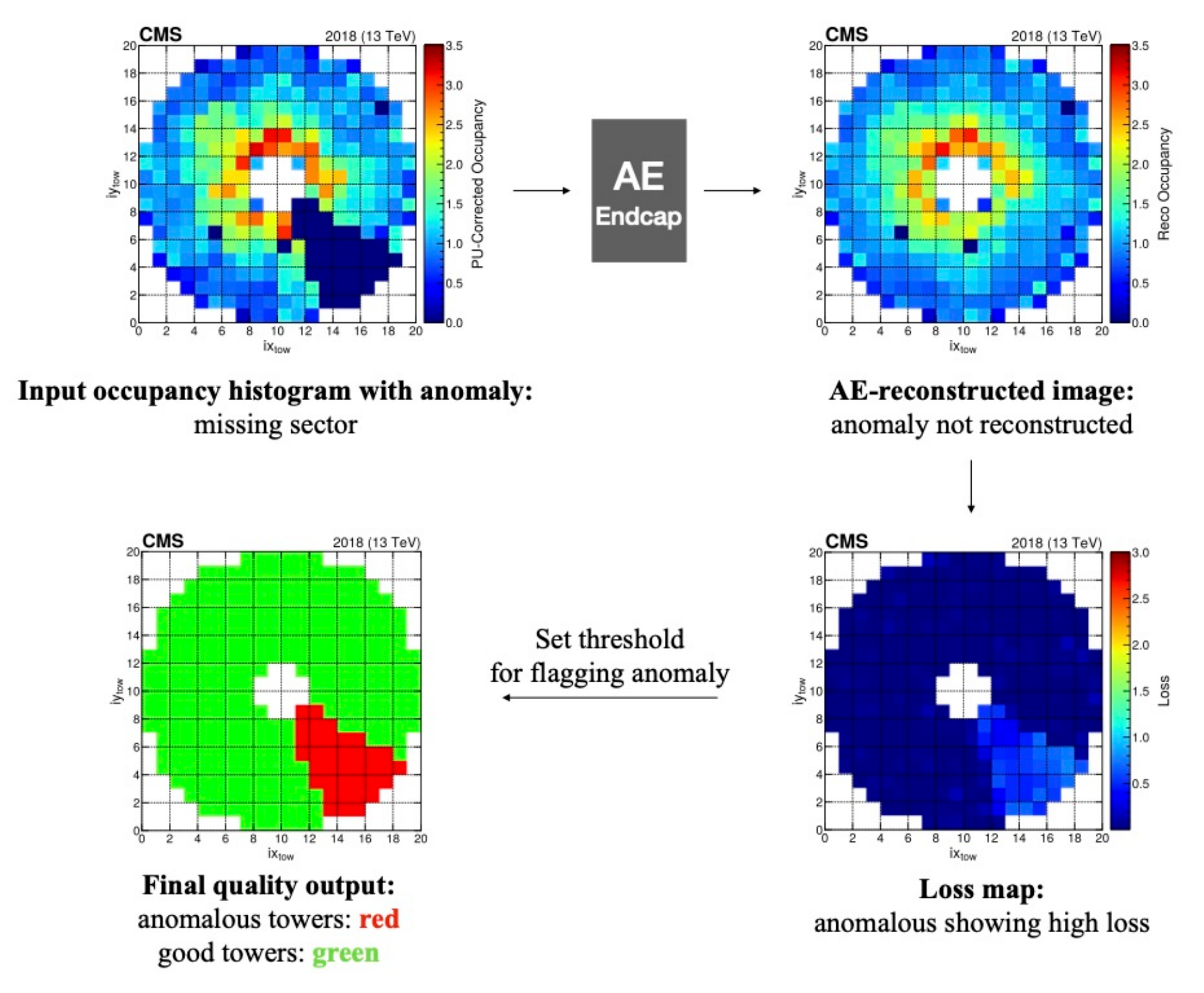}}
    \caption{Illustration of the AE-based anomaly detection strategy.}
    \label{fig:AE_strategy}
\end{figure}

Figure~\ref{fig:AE_strategy} illustrates the anomaly detection strategy for the ECAL DQM using endcap images as an example. The input occupancy image (top left) is fed to the AE, which outputs a reconstructed image (top right). Then the Mean Squared Error on each tower is calculated and plotted as a 2D loss map in the same coordinates ($i\eta_{tow}-i\phi_{tow}$ for the barrel and $ix_{tow}-iy_{tow}$ for the endcaps). As shown in the bottom-right panel, the anomalous region is highlighted with the loss higher than the rest of the image. After applying some post-processing steps explained in Sections~\ref{sec:spatialcorr} and \ref{sec:timecorr}, a threshold to flag the anomaly is calculated based on the anomalous loss values. The threshold is applied to the post-processed loss map to create a quality plot (bottom left), where towers with the loss above the threshold are tagged as anomalous and are shown in red. Towers with the loss below the threshold are identified as good and are shown in green. The final quality plot can be easily and quickly interpreted by a DQM shifter.

\section{Training, Evaluation, and Post-processing }
\label{sec:ML_part}

\subsection{Training and Validation}
\label{sec:train}
The available dataset consists of 100\,000 good images processed offline with each image corresponding to a single LS. 
This dataset is split into a training and a validation set with a ratio of 9:1. In addition to the validation set with good images, another validation set is obtained that comprise the same good images but with ``fake'' anomalies introduced.
Three kinds of anomalies are explored: 
\begin{itemize}
    \item Missing supermodule/sector: Entire barrel supermodules and endcap sectors are randomly set to have zero occupancy values in each LS. 
     \item Single zero occupancy tower: A single tower is set to have zero occupancy at random in each LS. Such low-occupancy single towers are usually harder to detect.
    \item Single hot tower: 
    A single tower is set to be ``hot'', or having higher-than-nominal occupancy.  
    For a tower with 25 crystals and 500 events per LS, the average occupancy is of the order:
    
    \begin{equation}
     \mathrm{occupancy} =  25 \times 500 \times f,
    \label{eq:occ-hot}
    \end{equation}
    
    where \textit{f} is the frequency of the readout.  For the barrel, nominal $f$ ranges from 0.01 to 0.03, while for the endcaps it ranges from 0.02 to 0.05. For $f=1$, the readout is said to be in ``full-readout''. Hot towers with $f=1$ are easier to detect as their values stand out clearly from the nominal value.
    Thus for the validation, a more challenging borderline scenario is targeted with $f=0.1$ for the barrel and $f=0.2$ for the endcaps. The readout frequency target is chosen to be higher for the endcaps, as the nominal occupancy values for the endcaps are larger in the higher $|\eta|$ region compared to the barrel.
\end{itemize}
These fake anomalies are used to derive thresholds on the loss maps for efficient anomaly tagging with the AE. While fake anomalies are representative examples of real anomalies that occur in the detector, the AE model is further tested on real anomalous data from the 2018 and 2022 runs as discussed in Section~\ref{sec:realanom}.

\subsection{The ECAL Spatial Response Correction}
\label{sec:spatialcorr}
\begin{figure}[tb]
\centering{
\subfloat[]{\includegraphics[width=0.46\textwidth]{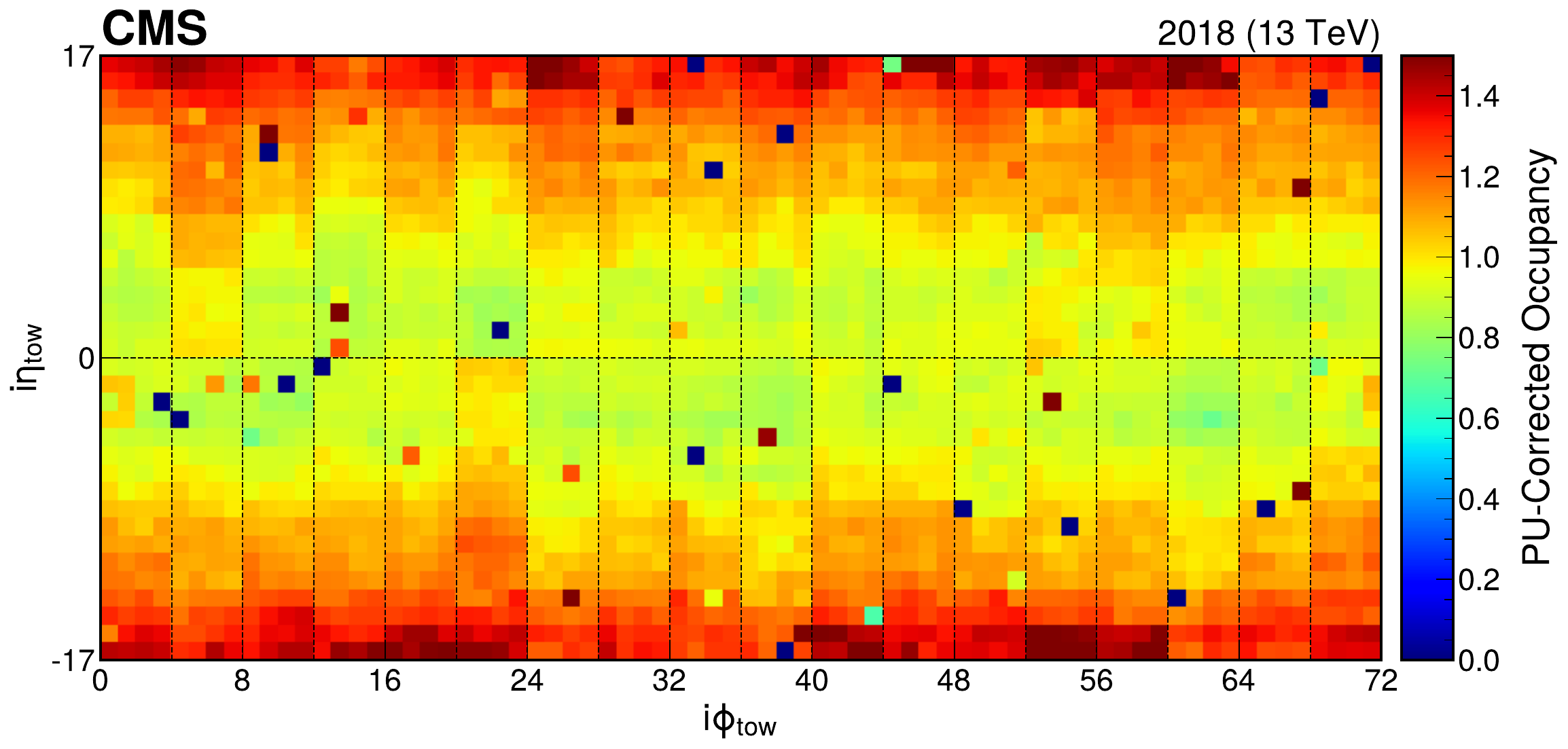}}
\subfloat[]{\includegraphics[width=0.25\textwidth]{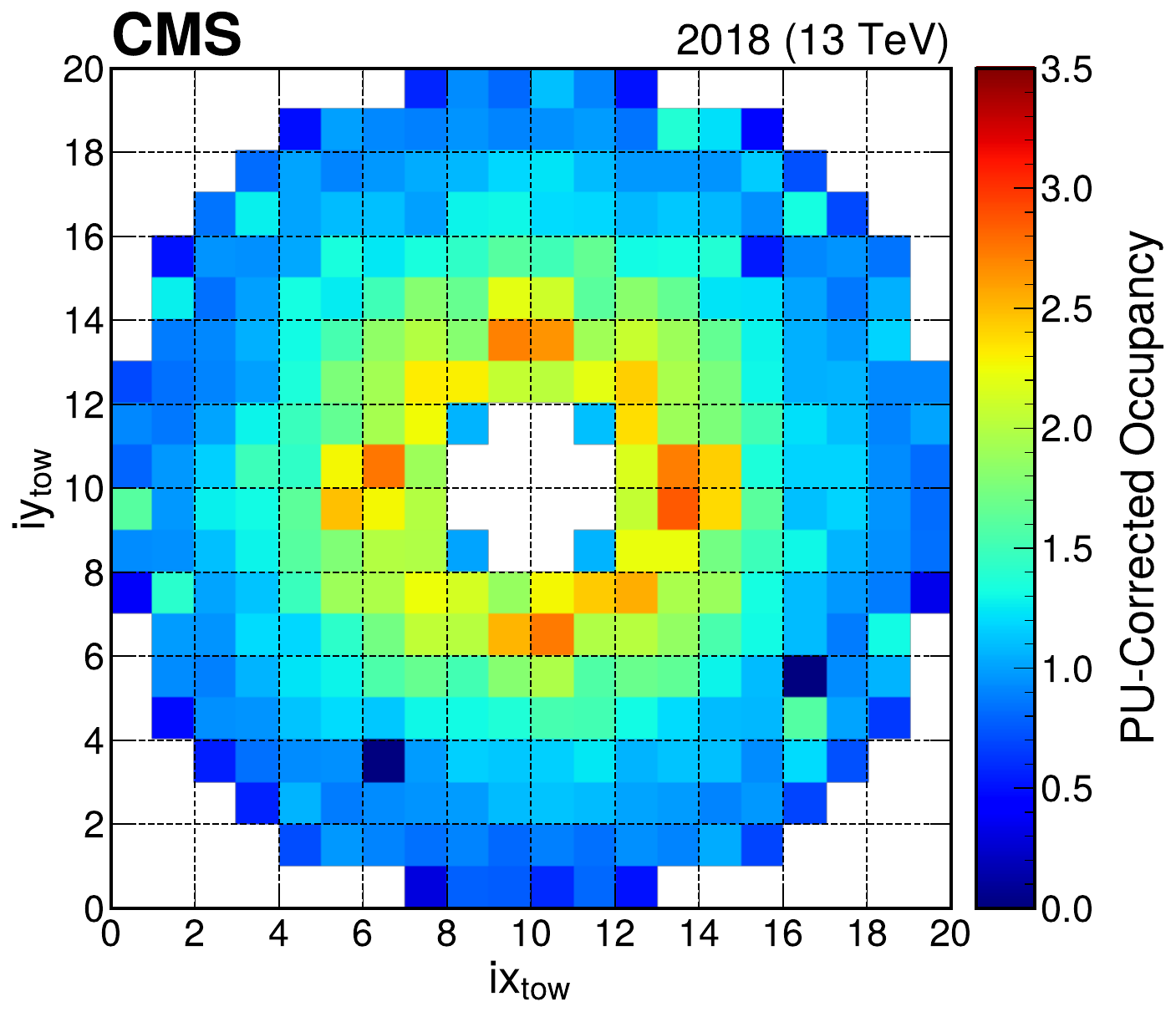}}
\subfloat[]{\includegraphics[width=0.25 \textwidth]{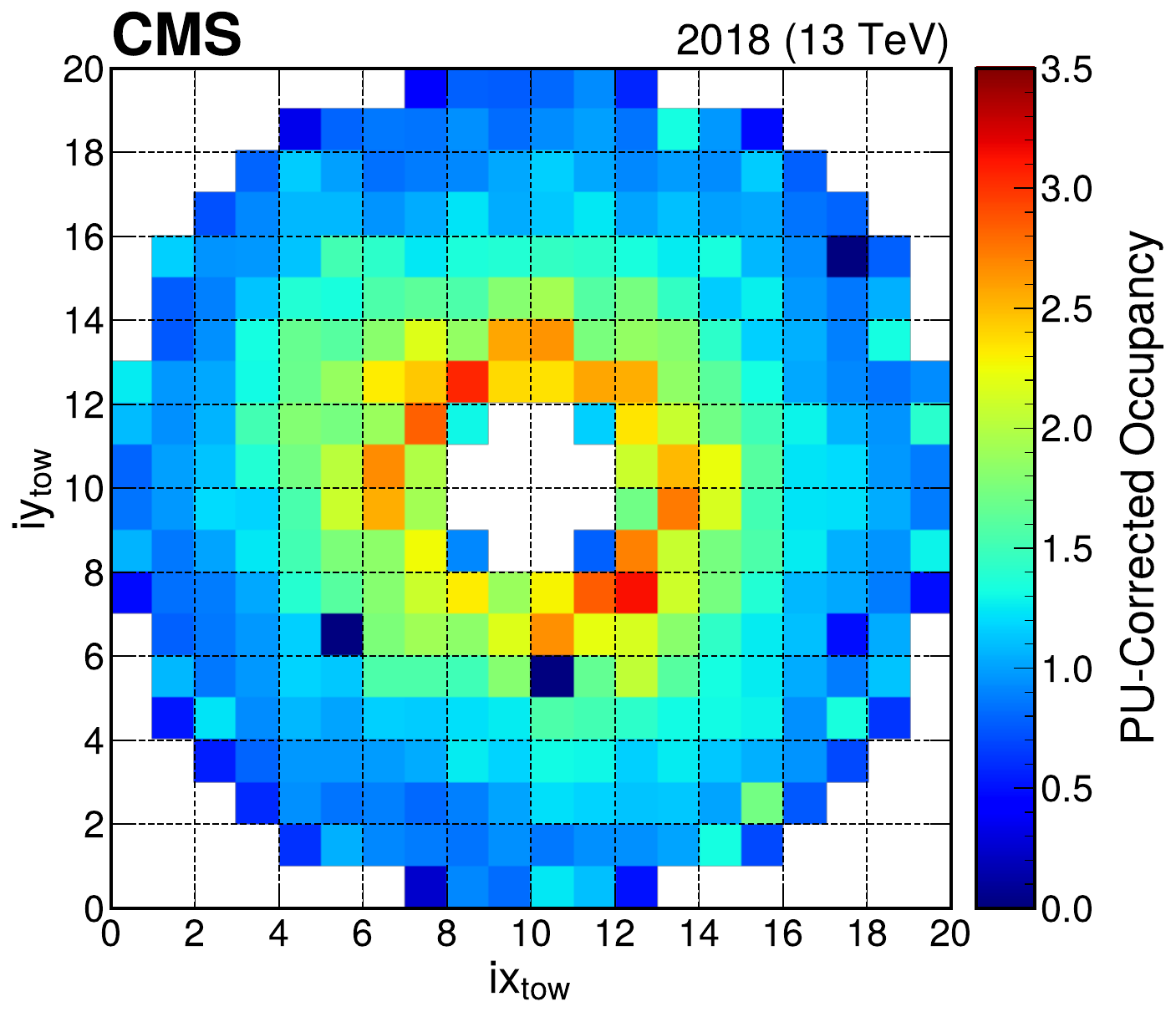}}
}
\caption{Average occupancy map for (a) the barrel, (b) EE$+$, and (c) EE$-$ from the 2018 dataset containing runs manually certified as good by the CMS data certification.}
\label{fig:avg}
\end{figure}

Since the multiplicity of particle production in a fixed rapidity interval is constant at a hadron collider, the number of particles per geometric interval increases for higher~$|\eta|$, which is related to rapidity. Due to this effect, it is observed that ECAL crystals in regions of high~$|\eta|$ exhibit higher occupancy than those of low~$|\eta|$ in both the barrel and the endcaps, as can be seen in Fig.~\ref{fig:avg}. 

\begin{figure}[btp]
\centering{
\subfloat[]{\includegraphics[width=0.47\textwidth]{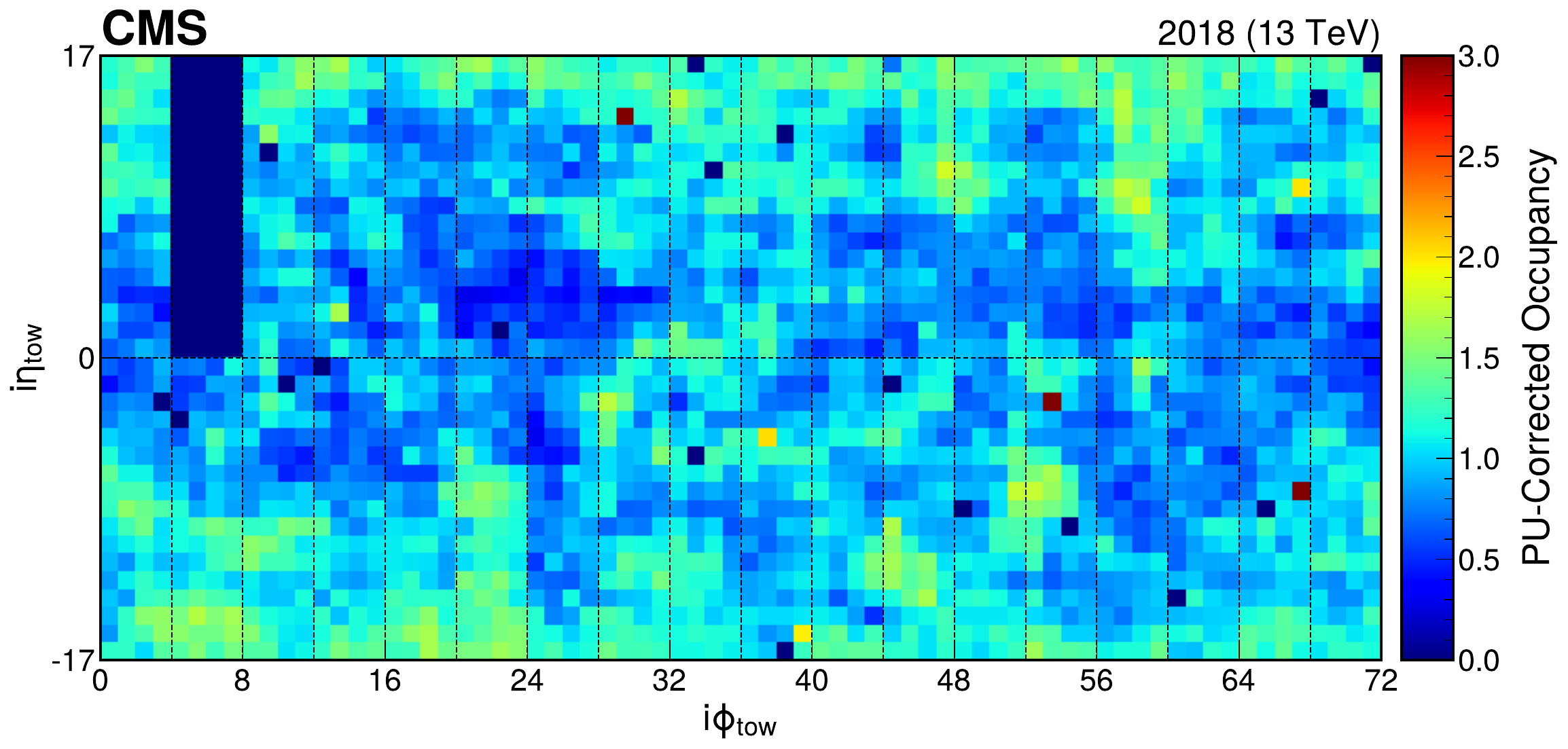}}
\hspace{0.2cm}
\subfloat[]{\includegraphics[width=0.47\textwidth]{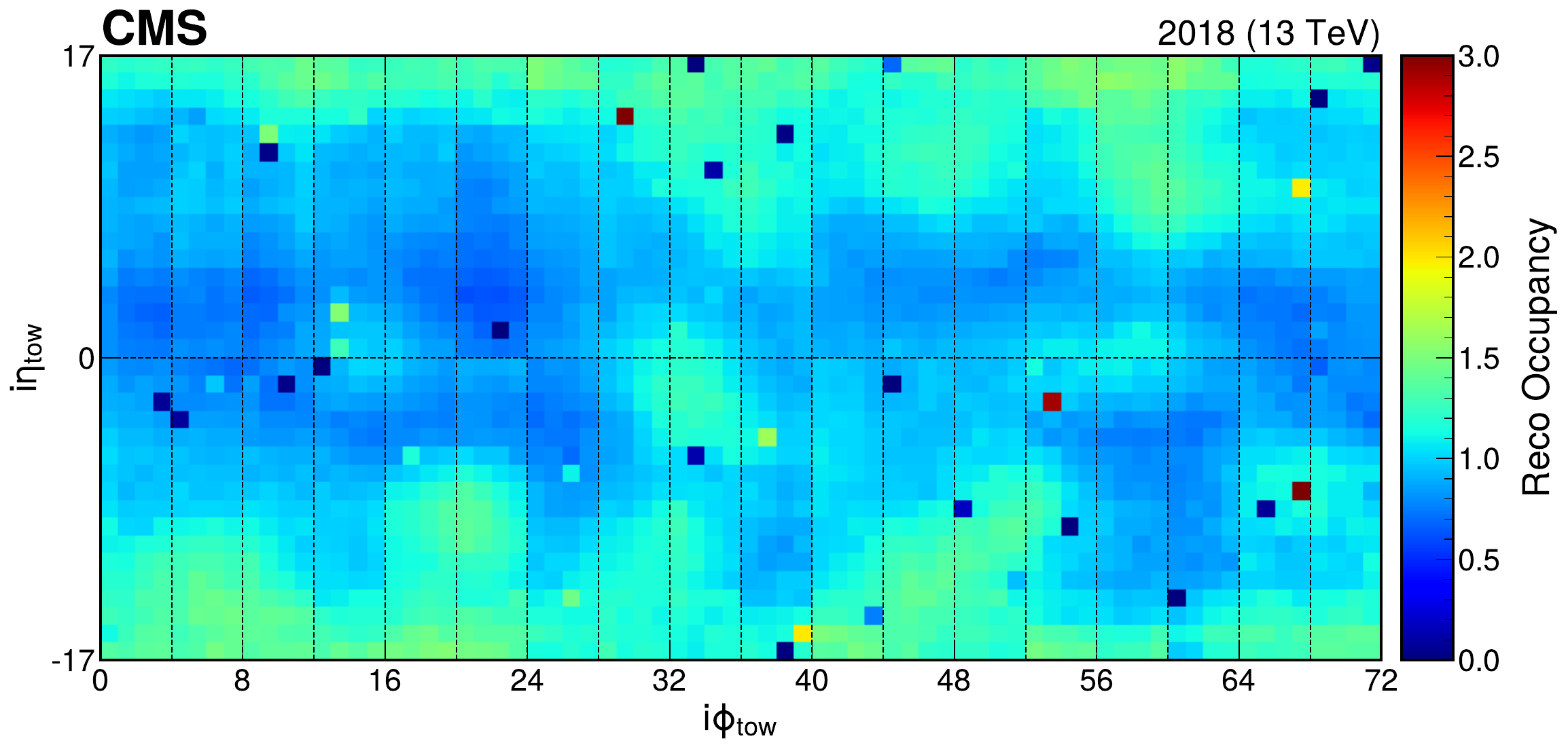}}
\\
\subfloat[]{\includegraphics[width=0.47\textwidth]{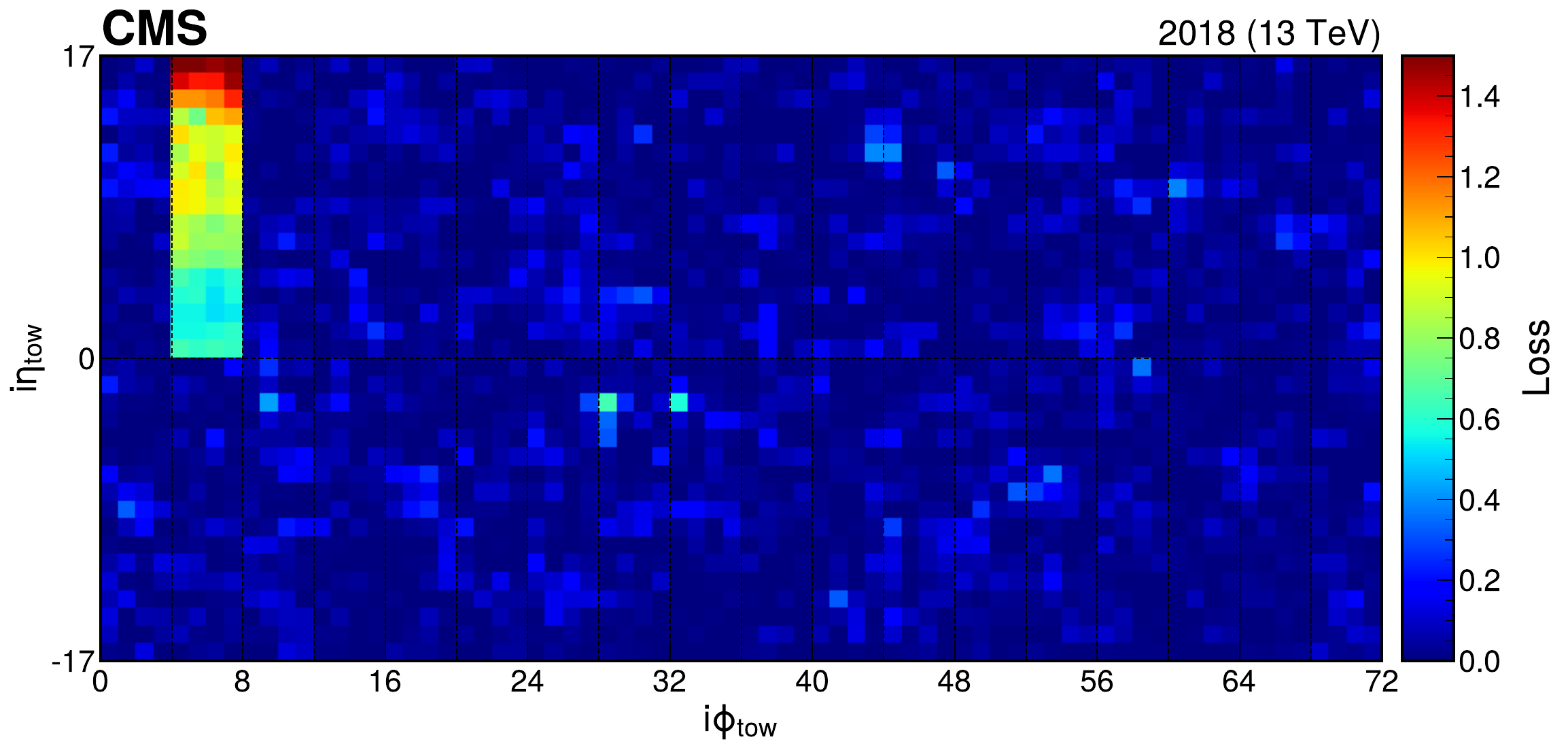}}
\hspace{0.2cm}
\subfloat[]{\includegraphics[width=0.47\textwidth]{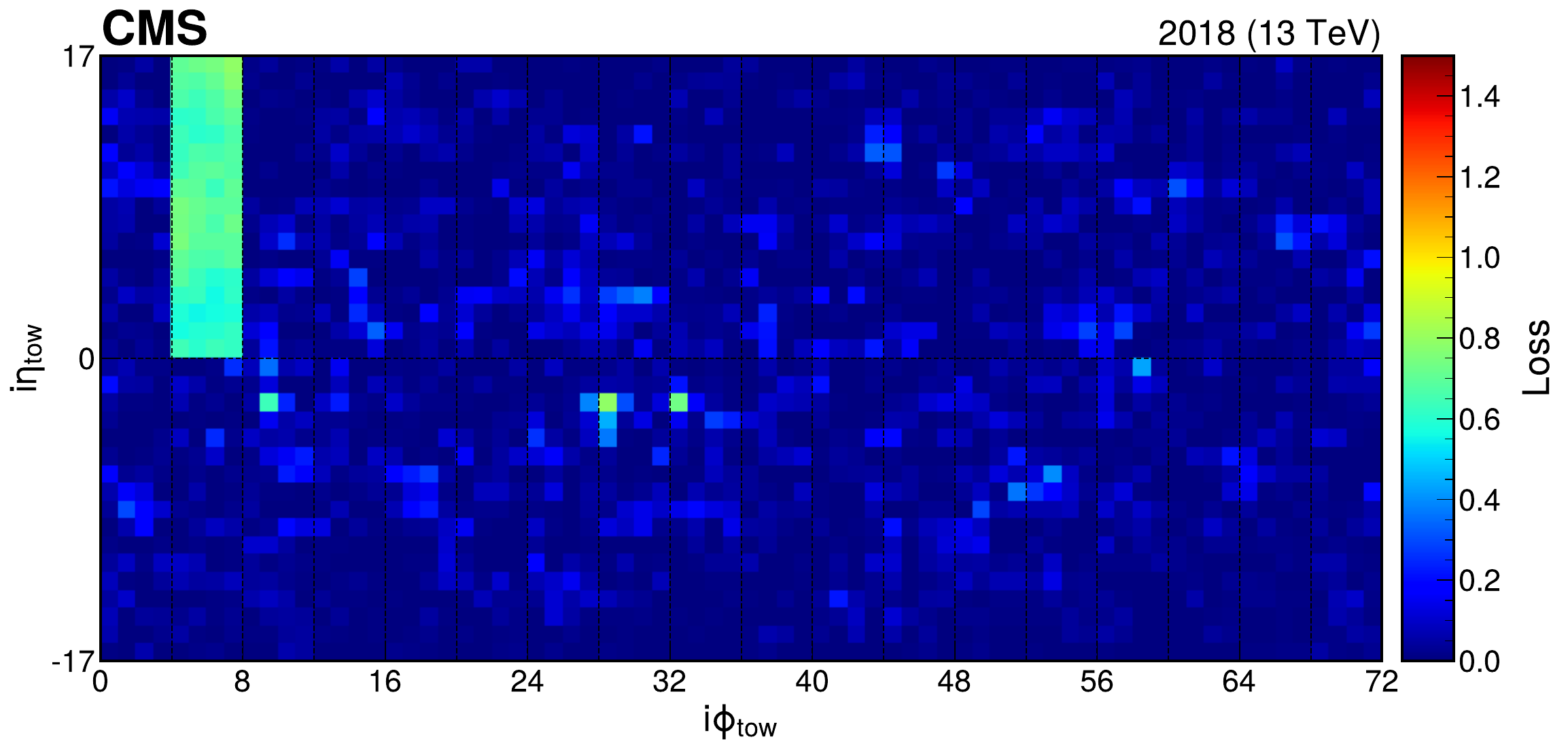}}
}
\caption{(a) Occupancy map with a missing supermodule in the barrel. (b) AE-reconstructed occupancy map. (c) Loss map showing the missing supermodule, indicating higher loss at high $|\eta|$ owing to differences in the detector response. (d) Loss map after the spatial correction is applied. }
\label{fig:SM}
\end{figure}

This difference in detector response is also visible in the AE loss map for specific anomalies as illustrated in Fig.~\ref{fig:SM} with a missing supermodule.
Figure~\ref{fig:SM}(a) shows the PU-corrected occupancy map with one supermodule having zero occupancy, and Fig.~\ref{fig:SM}(b) reflects the corresponding AE-reconstructed output where the AE fails to reconstruct the anomaly. 
Figure~\ref{fig:SM}(c) is the tower-level loss map calculated between the input and output, exhibiting high loss in the anomalous supermodule region; the towers at the highest $|\eta|$ tend to have a higher loss than those at lower $|\eta|$ due to the higher average occupancy in these regions. To mitigate this effect and obtain uniform loss in the anomalous region, 
the loss is normalized by the average occupancy shown in Fig.~\ref{fig:avg}(a). After this ``spatial response correction'', flat loss is observed in the anomalous region as seen in Fig.~\ref{fig:SM}(d), where all towers in the supermodule are interpreted as equally anomalous.

\subsection{Time Correction}
\label{sec:timecorr}
\begin{figure}[tbp]
\centering{
\includegraphics[width=1.0\textwidth]{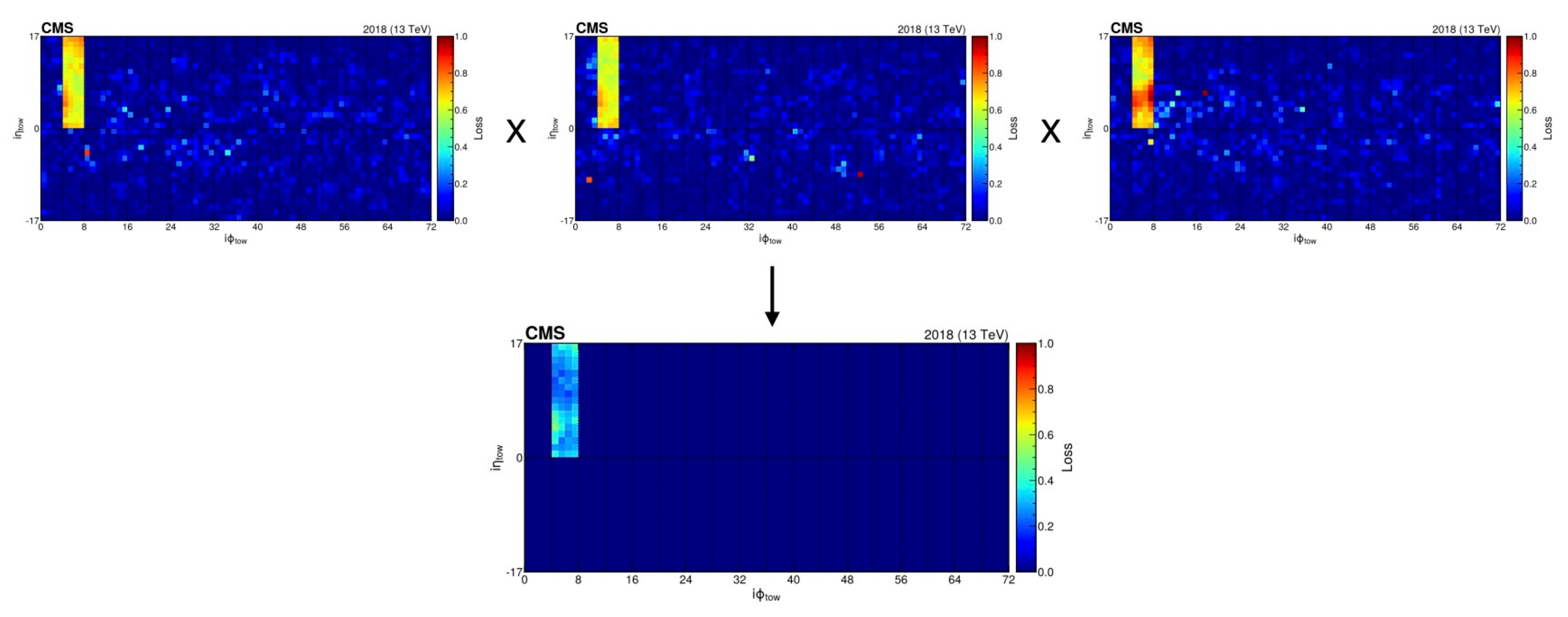} }
\caption{Time correction strategy: multiplication of loss maps from three consecutive LSs in the barrel (top) to obtain a final loss map (bottom), where only the real anomaly of the missing supermodule is highlighted with high loss, while random towers with false positives are suppressed.}
\label{fig:TimeMult}
\end{figure}

Real anomalies persist with time in consecutive LSs, while random fluctuations average out. A correction is implemented to exploit the time-dependent nature of the anomalies in the detector, named ``time correction'', which brings a significant improvement in the AE performance.
Figure~\ref{fig:TimeMult} shows the time correction strategy that is applied. 
Spatially corrected loss maps from three consecutive LSs (top panel) are multiplied together at a tower level. 
The resulting time-multiplied loss map at the bottom shows that the persistent anomaly of the missing supermodule is enhanced and random fluctuations visible in the loss maps from each LS are suppressed, reducing false positives.
It is observed that multiplication rather than averaging is a better strategy for enhancing and suppressing the resulting loss values. Multiplication results in the low loss of good towers being smaller and the high loss of bad towers being larger, widening the gap between both and thus enhancing the distinction between both scenarios.

Given the duration of $\sim$\,23 seconds for each LS with 500 events, time correction with three LSs yields a latency of approximately one minute. Including more LSs for the time correction is shown to bring no further significant reduction in the FDR in the AE performance during offline validation. Thus, one minute of latency is chosen to be an optimal trade-off for the time correction.
For the online deployment where the LSs contain different number of events, however, this is changed to six LSs as discussed in Section~\ref{sec:deploy}.

\subsection{Anomaly Tagging Threshold and Performance Metric}
\label{sec:FDR}
\begin{figure}[tb]
\centering{
\includegraphics[width=0.55\textwidth]{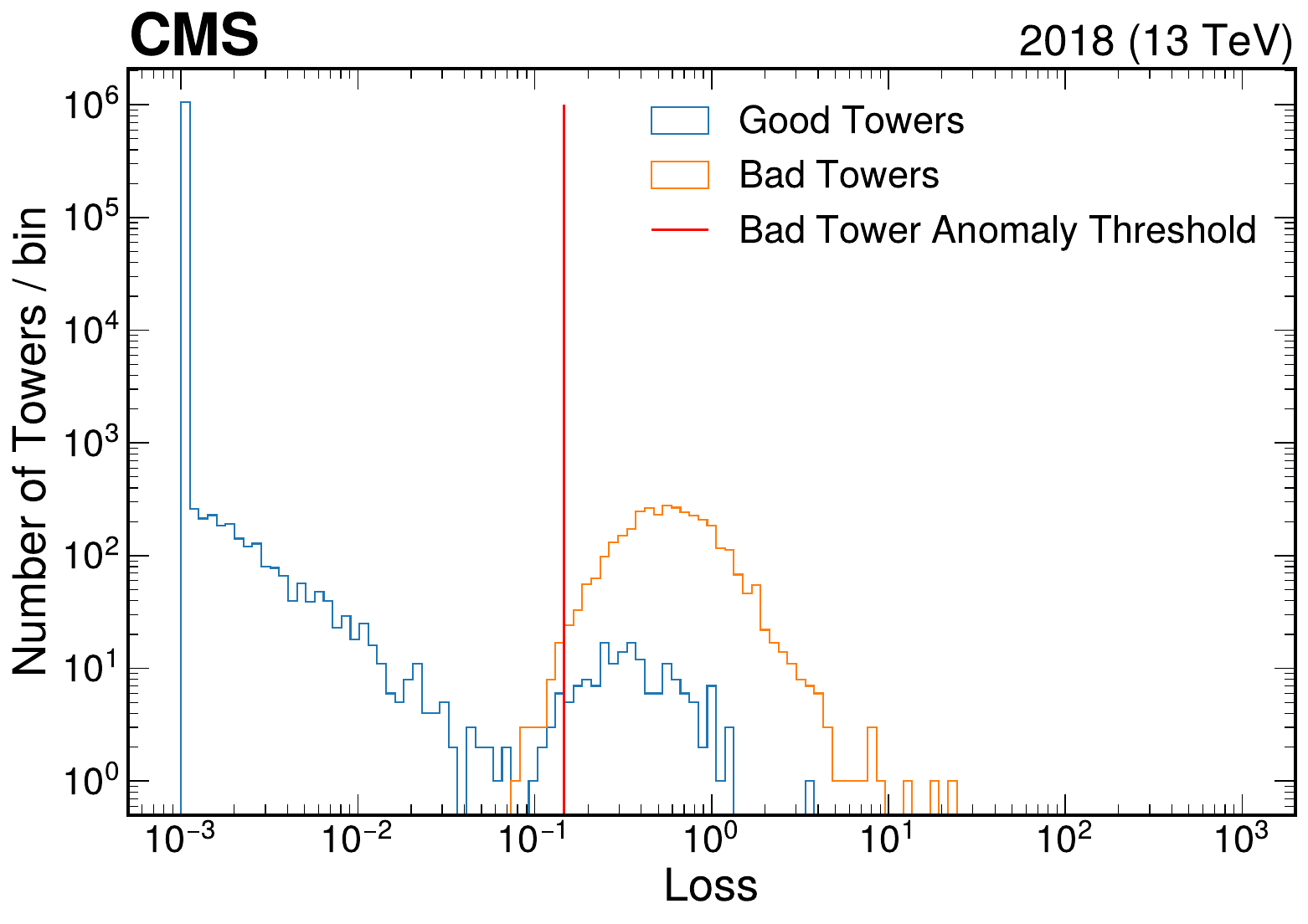} }
\caption{Loss distribution for the zero occupancy tower scenario after spatial and time correction for EE$-$. The anomaly threshold is set as the lower 1\% of the zero occupancy tower loss values. }
\label{fig:loss_thres}
\end{figure}

The goal of the ML-based DQM system is to maximize the anomaly detection efficiency while minimizing the number of false positives. 
An anomaly is tagged using a threshold obtained from a validation set with fake anomalies. The threshold on the final post-processed loss map is chosen such that the loss values of 99\% of anomalous towers are above the threshold as illustrated in Fig.~\ref{fig:loss_thres}, which shows the loss distribution from a zero occupancy tower scenario. The bump in the tail of the good tower loss distribution is discussed in Section~\ref{sec:fake}.

To assess the performance of the AE network, the False Discovery Rate~(FDR) is used as a metric and is defined as:

\begin{equation*}
    \textrm{{FDR}} = \frac{\textrm{{Number of good towers above $\epsilon$}}}{\textrm{{Number of good and bad towers above $\epsilon$}}}
\label{eq:FDR1}
\end{equation*}
where, $\epsilon$ is the threshold for anomaly detection.

The FDR value for 99\% anomaly detection represents the fraction of false detection in all anomalies detected, when using the threshold chosen to catch 99\% of the anomalies present in the dataset. In other words, the FDR is the ratio of good towers tagged as anomalous to all towers labeled as anomalous by the AE.
A lower FDR indicates better performance and fewer false alarms during data taking. The FDR is calculated for each anomaly scenario using the corresponding anomaly tagging threshold during validation.

\subsection{Comparison with Baseline Anomaly Detection Algorithm}
\label{sec:baseline}
A baseline study is carried out comparing the performance of the AE with a traditional cut-based approach. In this approach, baseline loss for a tower at a given~$(\phi, \eta)$ position is calculated for the barrel 
using the occupancy of the tower ($t_{\phi \eta}$) and the average occupancy of all towers within the same $\eta$-ring ($\langle t_\eta \rangle$) as:
\begin{equation}
    \textrm{Loss of tower}_{\phi \eta}  = | t_{\phi \eta} - \langle t_\eta \rangle|.
\label{eq:baseline_loss}
\end{equation}
This approach is equivalent to the way most anomaly thresholds are defined in the standard ECAL DQM.
A threshold is derived on this baseline loss 
using the same criteria as that for the AE, of achieving 99\% anomaly detection from the fake anomaly validation.
A similar baseline loss is not attempted for the endcaps, since the gradient of occupancy across the towers is much larger for the endcaps even within the same $\eta$-ring, and thus such baseline loss would not be a good method for detecting bad towers.

\section{Results}
\label{sec:Results}
\subsection{Testing on Fake Anomalies}
\label{sec:fake}

The performance of the AE-based DQM method is studied first on three distinct anomaly scenarios -- missing supermodule/sector, single zero occupancy tower, and single hot tower -- where artificial (fake) anomalies are added onto good images as outlined in Section~\ref{sec:train}.
Tables~\ref{tab:barrel} and \ref{tab:endcap} summarize the FDR values calculated with anomaly tagging thresholds determined for each scenario for 99\% anomaly detection. For the barrel where the baseline scenario is studied for comparison, it can be seen that the AE outperforms the baseline for all anomaly scenarios considered.
For both the barrel and the endcaps, the FDRs for the single zero occupancy tower scenario are observed to be always higher than that for the single hot tower case. 
This is because hot towers are in general easier to spot, as they stand out with much higher occupancy compared to neighboring towers of average occupancy. 

\begin{table}[tb]
\centering{
\caption{Summary of FDR using 99\% anomaly detection threshold for the ECAL barrel fake anomaly scenarios.}
\resizebox{0.8\textwidth}{!}{\begin{tabular}{|c|c|c|c|}
\hline
\multicolumn{1}{|l|}{} & \multicolumn{3}{c|}{FDR for 99\% anomaly detection}   \\ \hline
\multicolumn{1}{|l|}{}                                                    & Missing Supermodule & \multicolumn{1}{l|}{Zero Occupancy Tower} & \multicolumn{1}{l|}{Hot Tower} \\ \hline
\begin{tabular}[c]{@{}c@{}}Baseline \\ no correction\end{tabular} & 14\%      & 90\%                            & 5.2\%                          \\ \hline
\begin{tabular}[c]{@{}c@{}}Baseline after \\ time correction\end{tabular} & 5.9\%      & 80\%                            & $<$ 0.01\%                          \\ \hline
\begin{tabular}[c]{@{}c@{}}AE \\ no correction \end{tabular}      & 3.6\%      & 51\%                            & 2.8\%                          \\ \hline
\begin{tabular}[c]{@{}c@{}}AE after\\  spatial correction\end{tabular}      & 3.1\%      & 49\%                            & 2.9\%                          \\ \hline
\begin{tabular}[c]{@{}c@{}}AE after\\  spatial and \\ time corrections\end{tabular}       & 0.13\%     & 4.1\%                           & $<$ 0.01\%                          \\ \hline
\end{tabular}}
\label{tab:barrel}
}
\end{table}

\begin{table}[tb]
\centering{
\caption{Summary of FDR using 99\% anomaly detection threshold for fake anomaly scenarios in the endcaps.}
\resizebox{0.8\textwidth}{!}{\begin{tabular}{|c|c|c|c|c|c|c|}
\hline
\multicolumn{1}{|l|}{} & \multicolumn{6}{c|}{FDR for 99\% anomaly detection}   \\ \hline
\multicolumn{1}{|c|}{}                                    & \multicolumn{2}{c|}{Missing Sector} & \multicolumn{2}{c|}{Zero Occupancy Tower} & \multicolumn{2}{c|}{Hot Tower} \\ \hline
 & EE$+$ & EE$-$ & EE$+$ & EE$-$ & EE$+$ & EE$-$ \\ \hline
\begin{tabular}[c]{@{}c@{}}AE \\  no correction\end{tabular}      & 29\%      & 28\%                            & 86\%      & 86\%         & $<$ 0.01\%      & $<$ 0.01\%       \\ \hline
\begin{tabular}[c]{@{}c@{}}AE after\\  spatial correction\end{tabular}      & 1.8\%     & 2.2\%       & 11\%     & 14\%                      & 0.02\%               & 0.04\% \\
\hline
\begin{tabular}[c]{@{}c@{}}AE after\\  spatial and \\ time corrections\end{tabular}      & 0.06\%      & 0.18\%                            & 1.4\%      & 4.4\%     & $<$ 0.01\%      & $<$ 0.01\%     \\ \hline
\end{tabular}}
\label{tab:endcap}
}
\end{table}

The effect of each consecutive correction on the FDRs can be seen from the tables. 
The AE spatial correction reduces the FDRs in the missing supermodule/sector and the single zero occupancy tower scenarios, where the occupancy values are set to zero for the barrel/endcaps.
Without the correction, the loss values for the towers with zero occupancy anomalies are proportional to the towers' nominal occupancy, which indicates that the loss is biased to be larger in the higher~$|\eta|$ region (see e.g. Fig.~\ref{fig:SM}(c)). The corresponding loss map exhibits an effective gradient across the map, following that of the average occupancy map shown in Fig.~\ref{fig:avg}. It can be seen that the spatial correction has a greater effect for the endcaps than for the barrel, as the gradient in occupancy values across the towers is more pronounced for the endcaps.

In the case of the hot tower anomaly, the FDRs increase after the spatial correction. This is because the hot tower loss is biased to be higher in the opposite direction, towards the lower~$|\eta|$ region.
This leads to different effects of the spatial correction for different anomaly scenarios. For zero occupancy towers, spatial correction flattens out the gradient in the loss distribution and improves their detection. For the hot towers, the gradient is enhanced and the AE performance slightly drops.
However, this effect is mitigated by the time correction that greatly improves the FDRs for all anomaly scenarios, with excellent final performance scores for both the barrel and the endcaps. 

The remaining false positives, which show up as apparently ``good towers'' above the anomaly threshold after the time correction, are very likely to be actual anomalous towers that have been undetected so far in the dataset of good LSs. These towers show up with the higher loss in the tail of the good tower loss distribution (see Fig.~\ref{fig:loss_thres}) and contribute to overestimating the FDR.

\begin{figure}[h!]
\centering{
\includegraphics[width=0.42\textwidth]{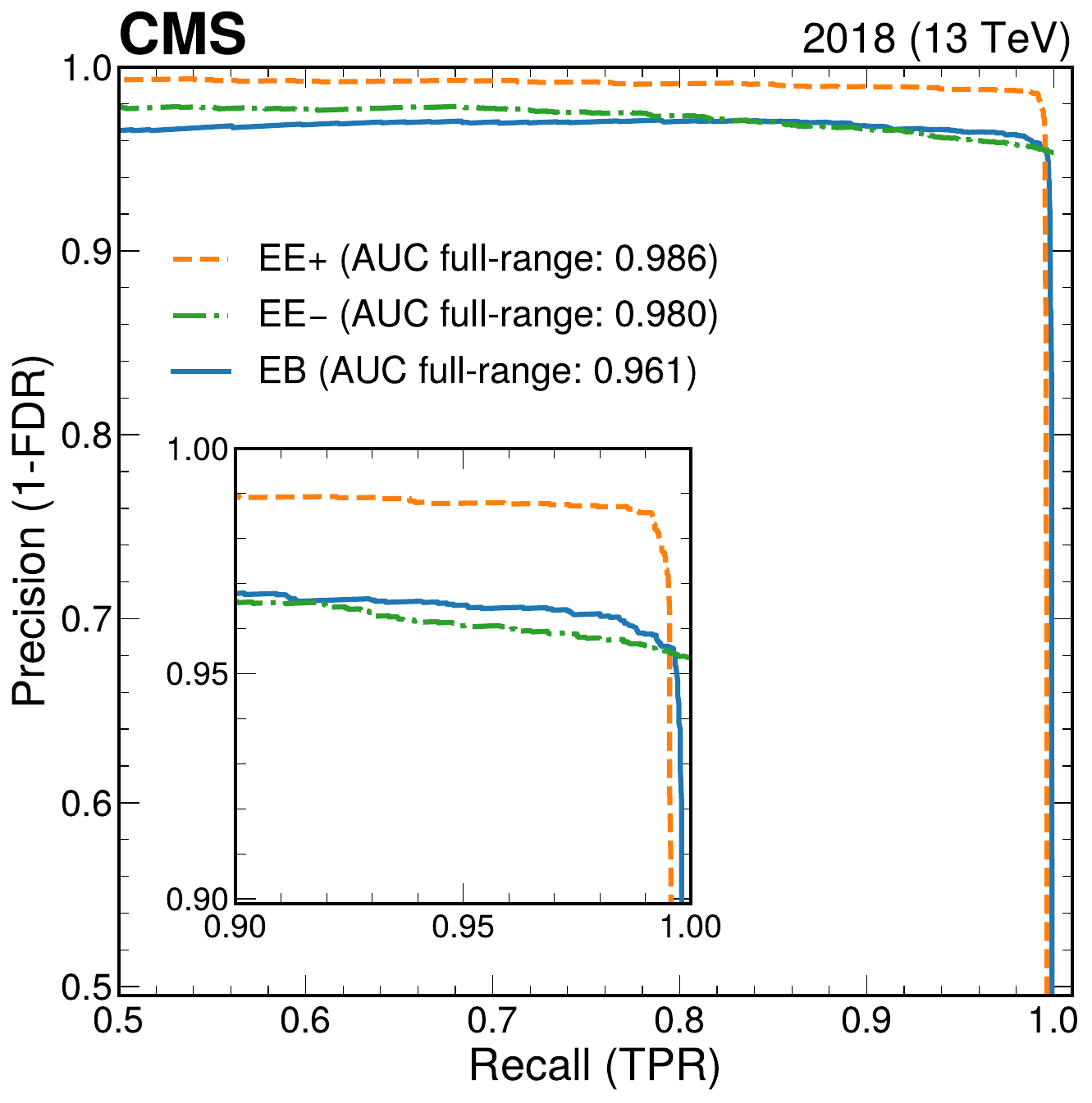} }
\caption{Precision (1-FDR) versus Recall (TPR) curve for the zero-occupancy scenario after spatial and time corrections are applied. The inlay plot shows a zoomed-in view. The full-range area under the curves (AUC) are also indicated.}
\label{fig:PR_curve}
\end{figure}

While 99\% anomaly tagging efficiency is chosen as the working point, FDR values at different working points can be seen from Fig.~\ref{fig:PR_curve} for the most challenging scenario of zero-occupancy tower after the spatial and time corrections are applied. The observed difference in the performance of the different parts of the detector is attributed to the different amounts of contamination of actual anomalous towers in the validation dataset. In both barrel and endcaps, high-precision high-recall of the AE-based anomaly detection is clearly displayed. 

\subsection{Testing on Real Anomalies}
\label{sec:realanom}
\begin{figure}[tbp]
\centering{
\subfloat[From a 2018 run: Input occupancy map with an error in the supermodule~EB$-$03]{\includegraphics[width=0.47\textwidth]{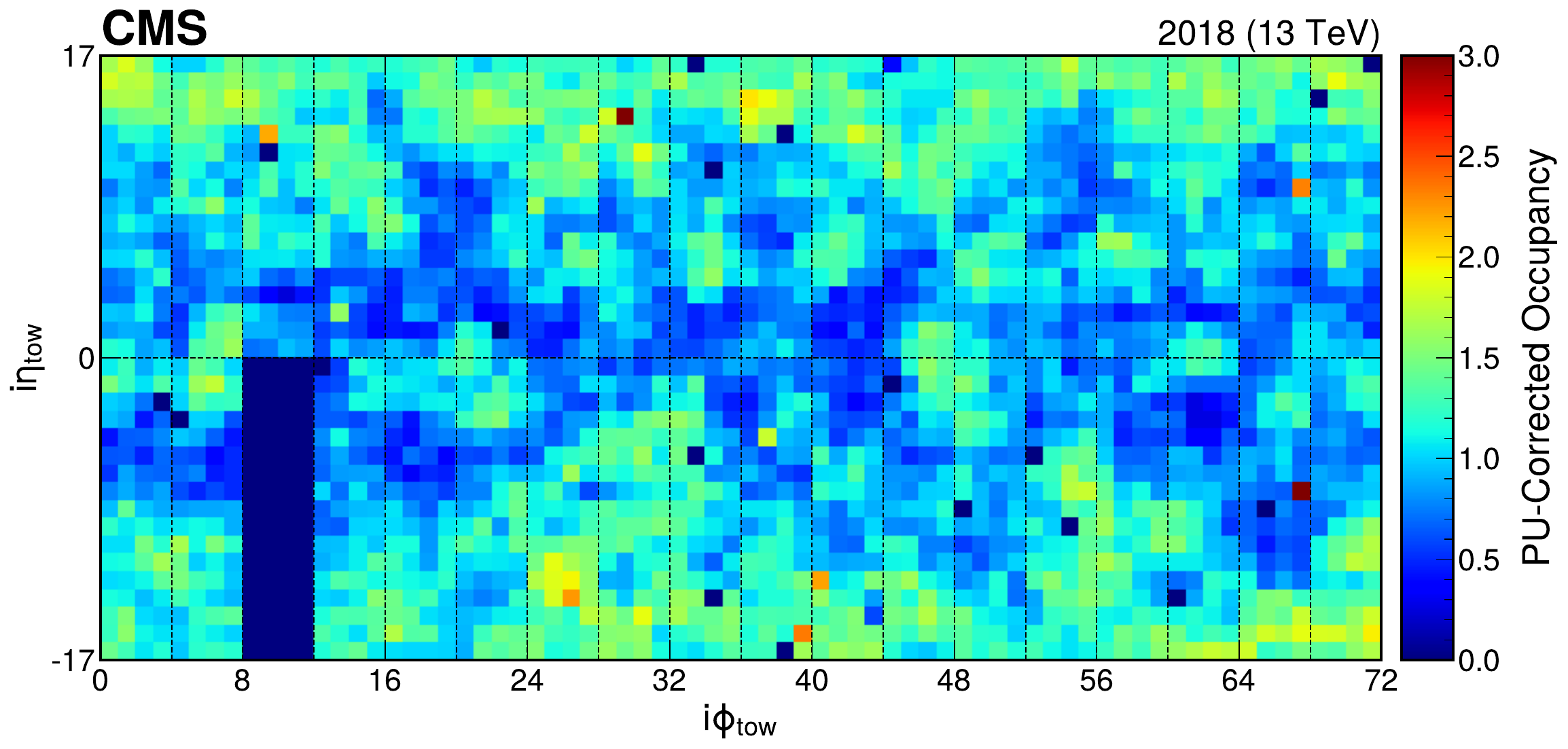}}
\subfloat[From a 2018 run: Final AE quality plot]{\includegraphics[width=0.42\textwidth]{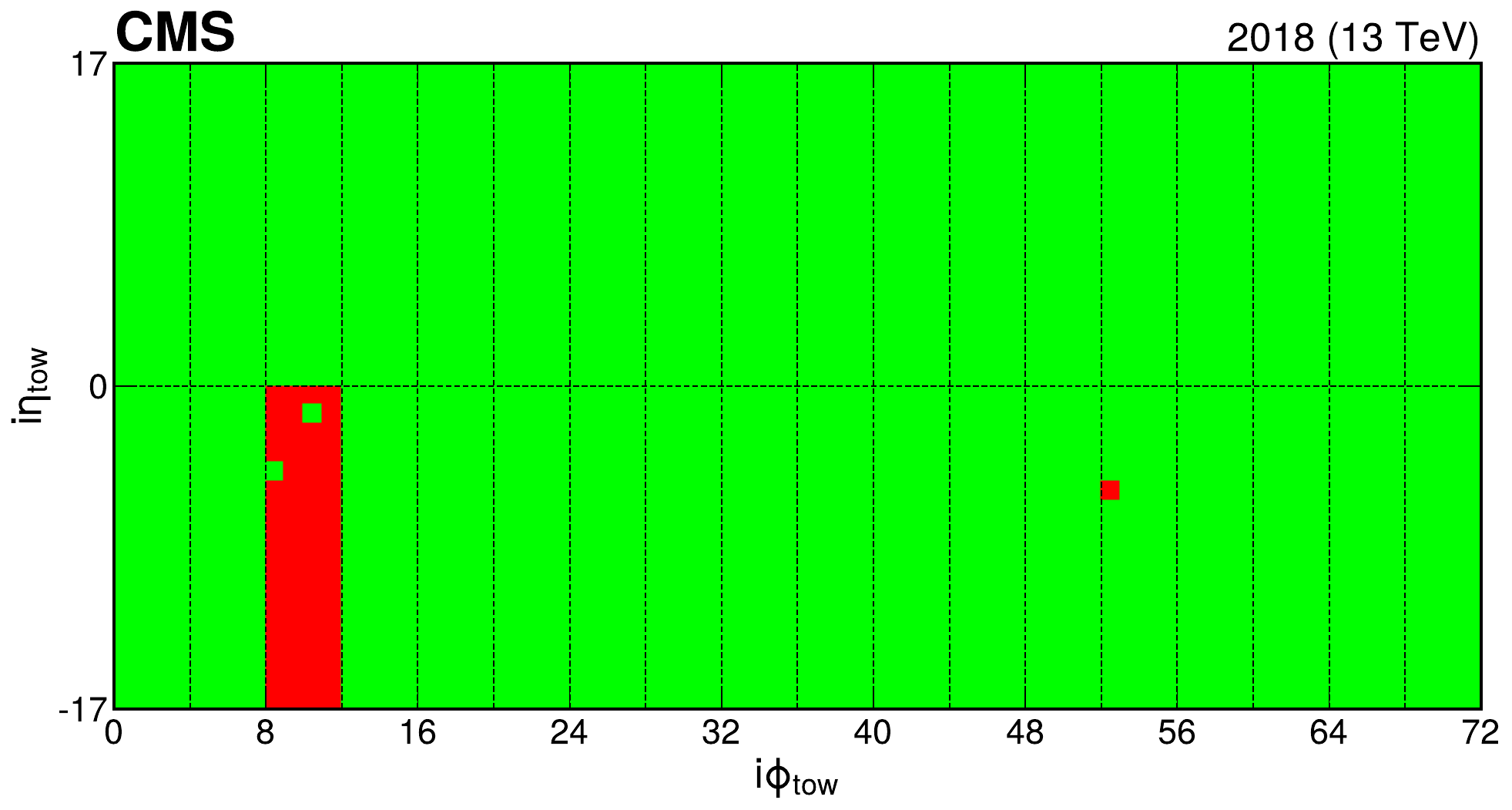}}
\\
\subfloat[From a 2018 run: Input occupancy plot with a group of hot towers]{\includegraphics[width=0.47\textwidth]{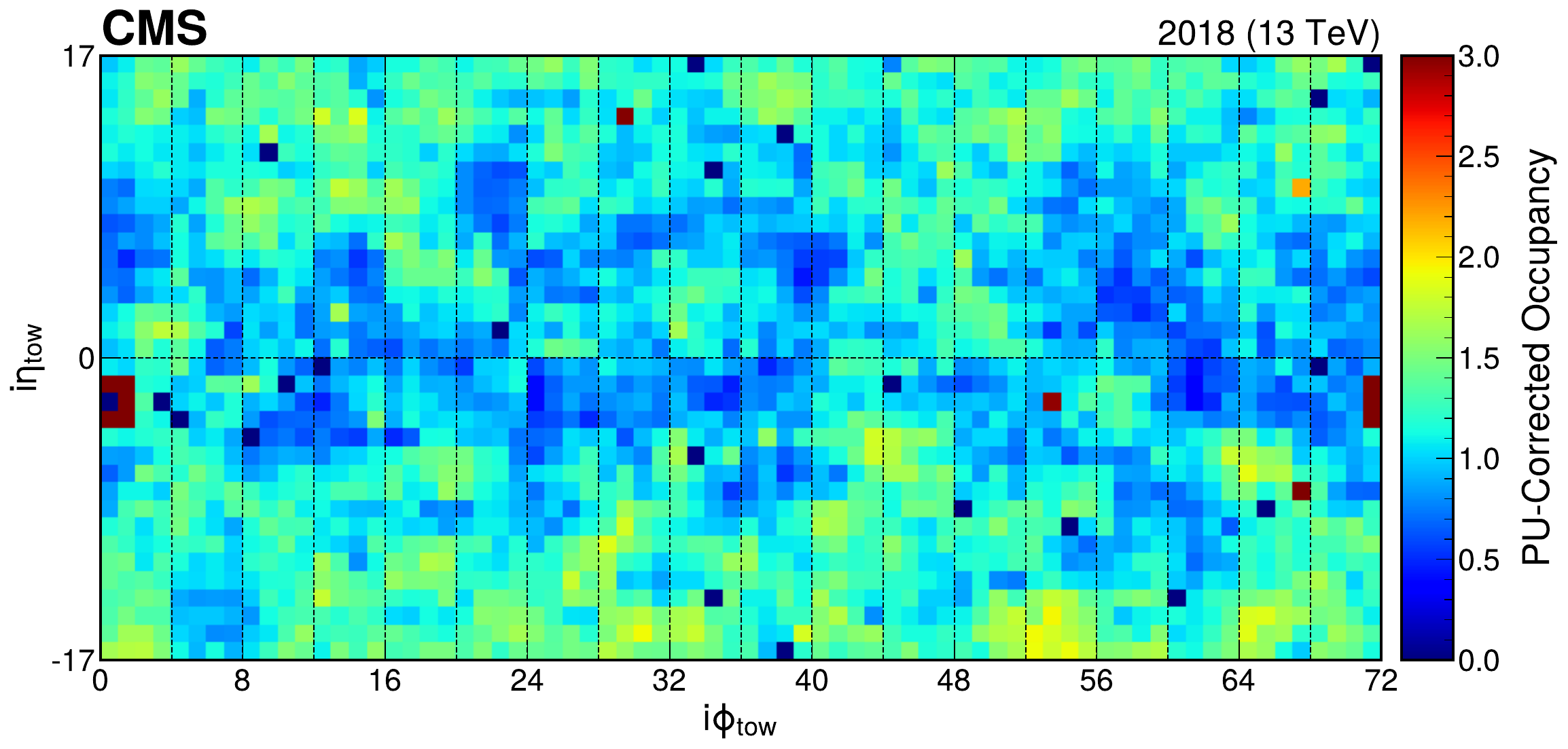}}
\subfloat[From a 2018 run: Final AE quality plot]{\includegraphics[width=0.42\textwidth]{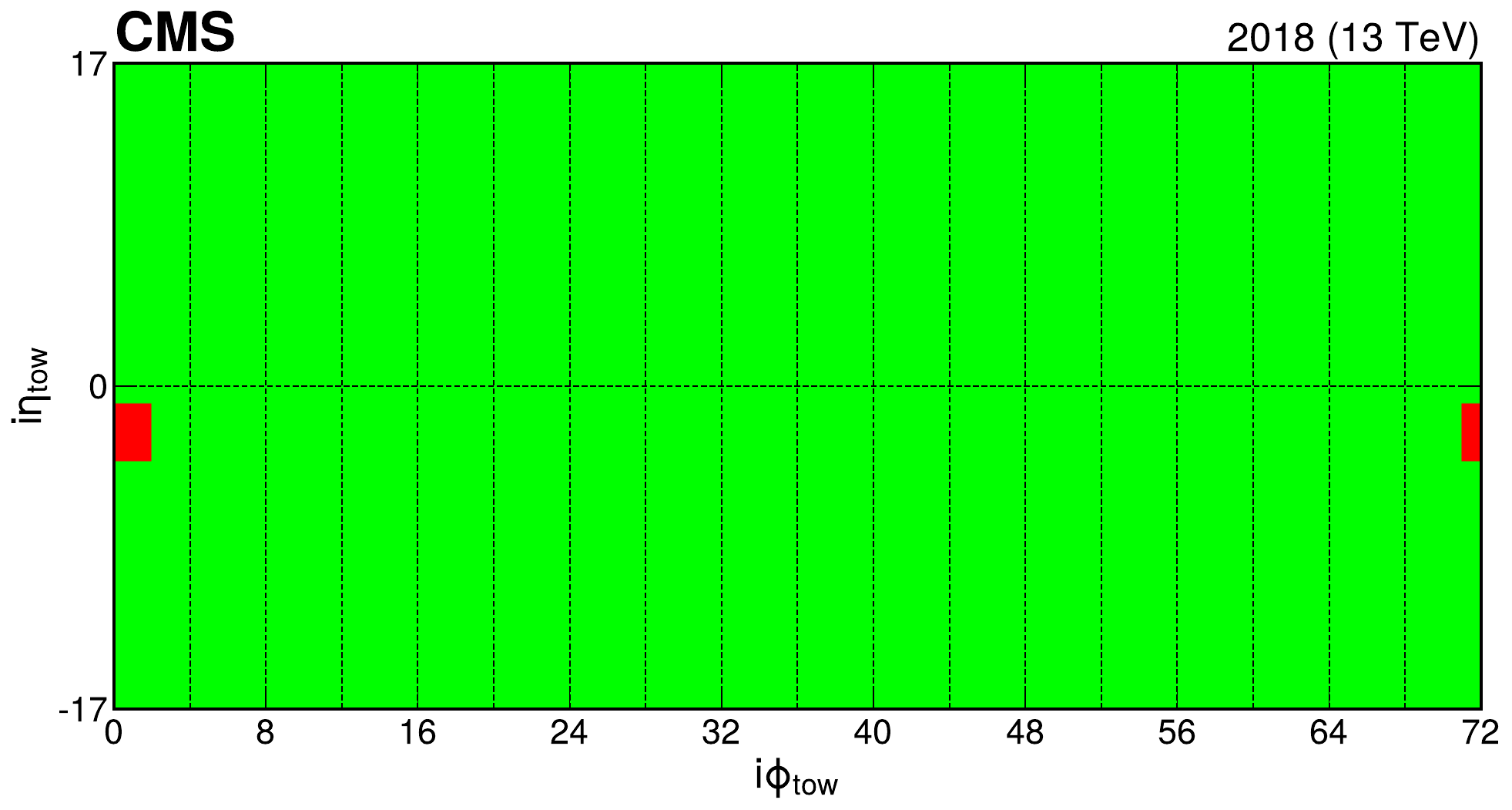}}
}
\caption{Input occupancy images with real anomalies and corresponding AE quality plots for the barrel.}
\label{fig:realanom_EB}
\end{figure}

\begin{figure}[tb]
\centering{
\subfloat[From a 2018 run: Input occupancy plot with more than half of the EE$+$ turned off]{\includegraphics[width=0.37\textwidth]{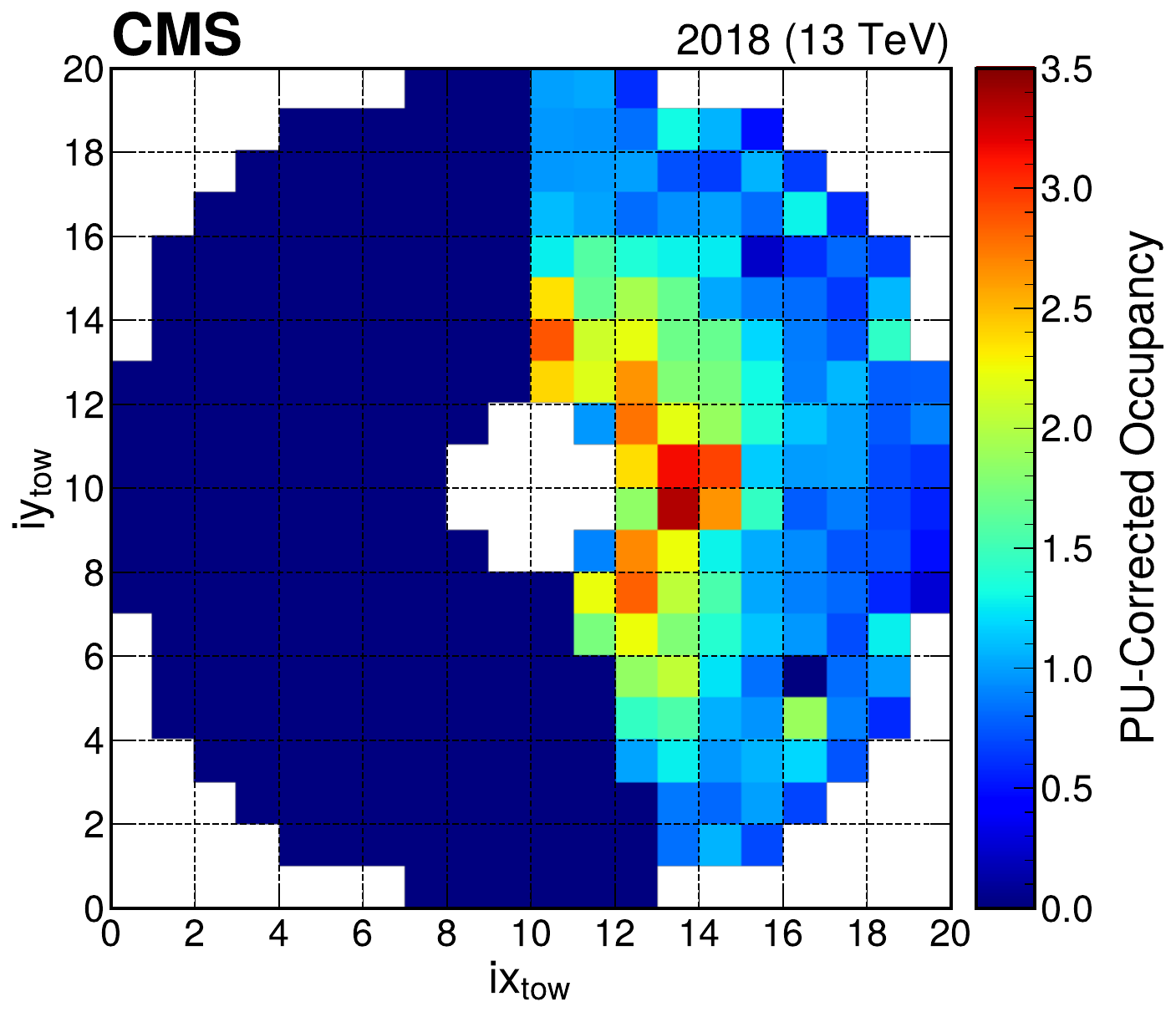}}
\hspace{0.5cm}
\subfloat[From a 2018 run: Final AE quality plot]{\includegraphics[width=0.31\textwidth]{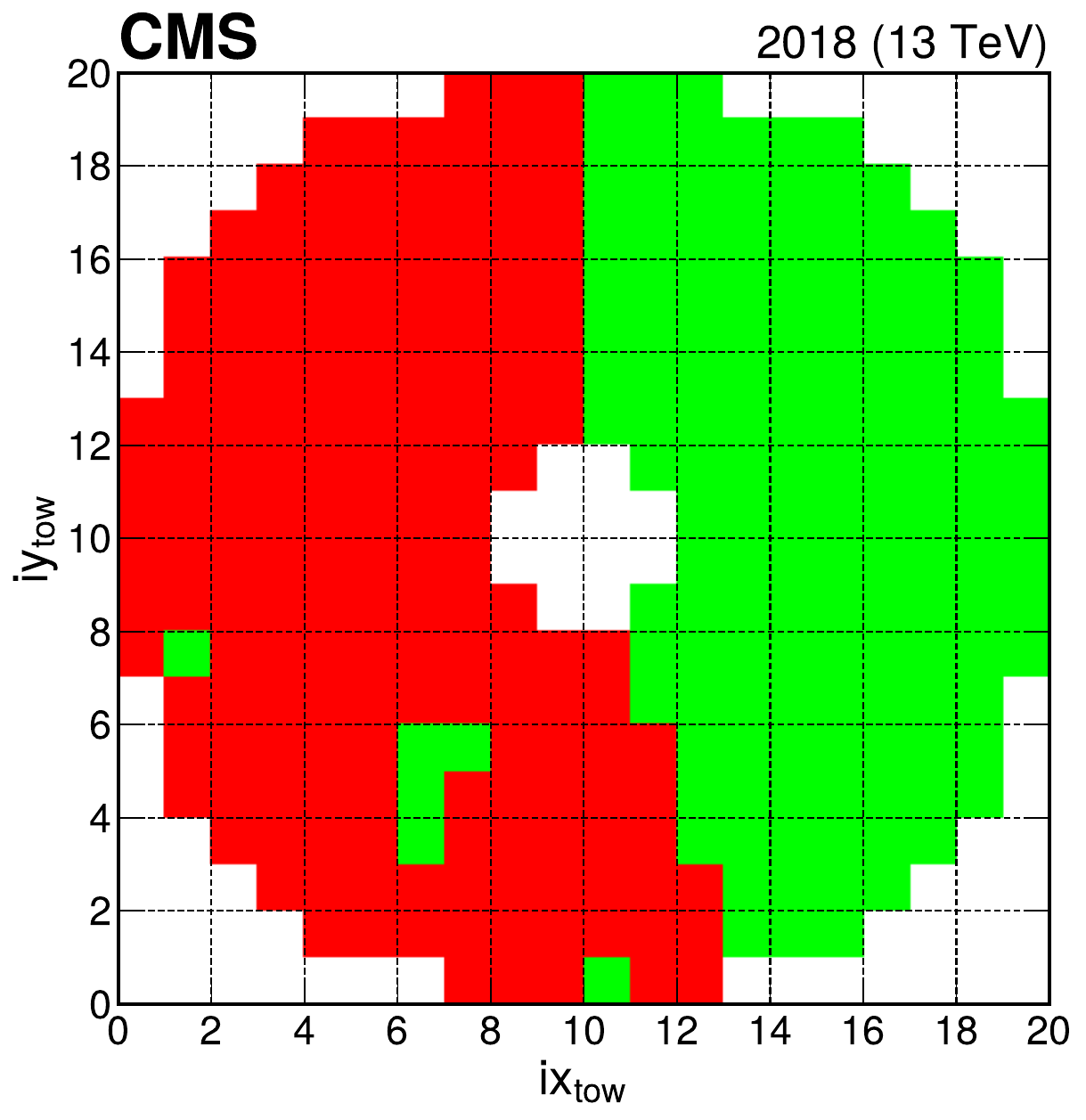}}
\\
\subfloat[From a 2022 run: Input occupancy plot with a region of towers turned off in EE$+$]{\includegraphics[ width=0.37\textwidth]{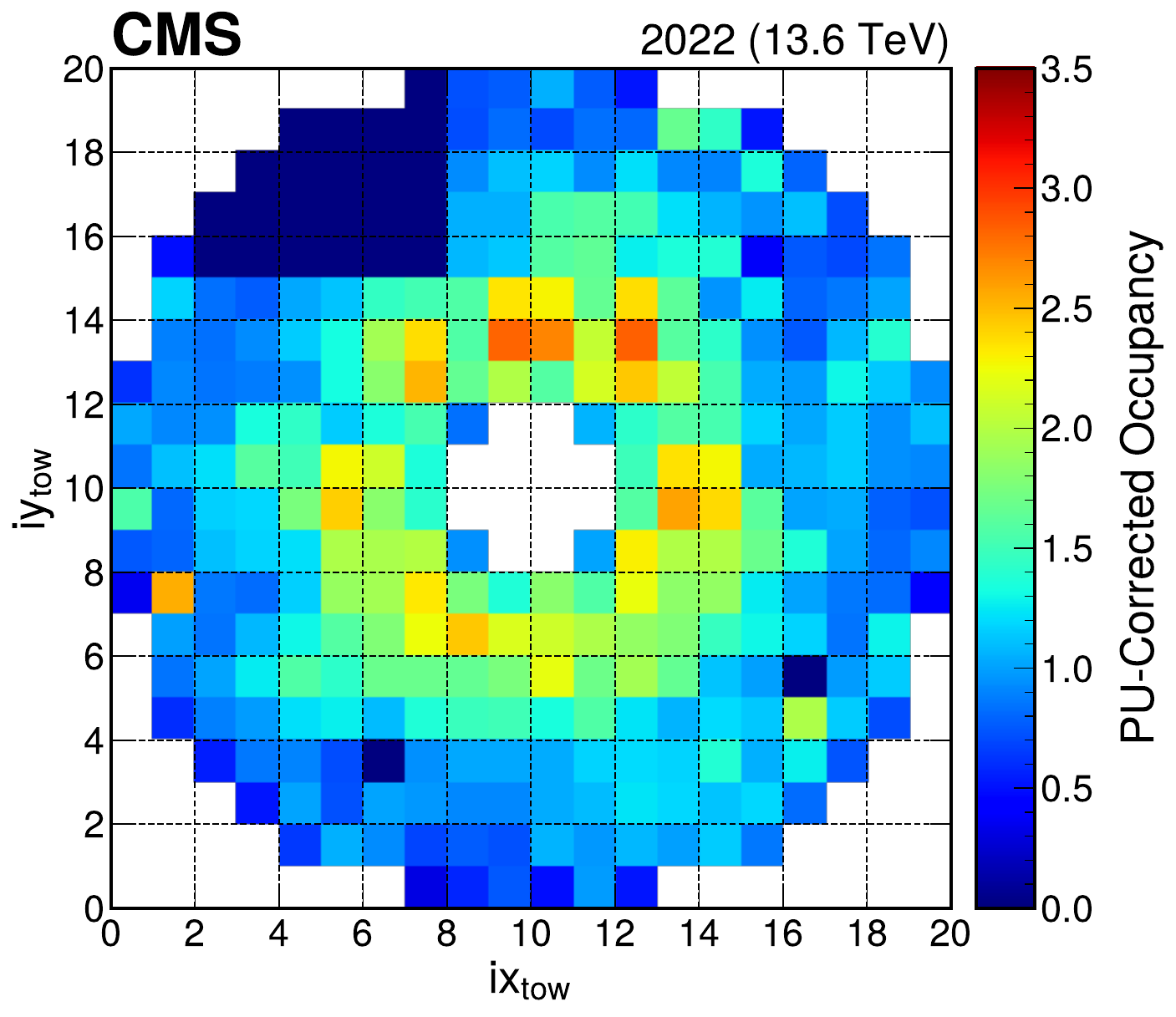}}
\hspace{0.5cm}
\subfloat[From a 2022 run: Final AE quality plot]{\includegraphics[ width=0.31\textwidth]{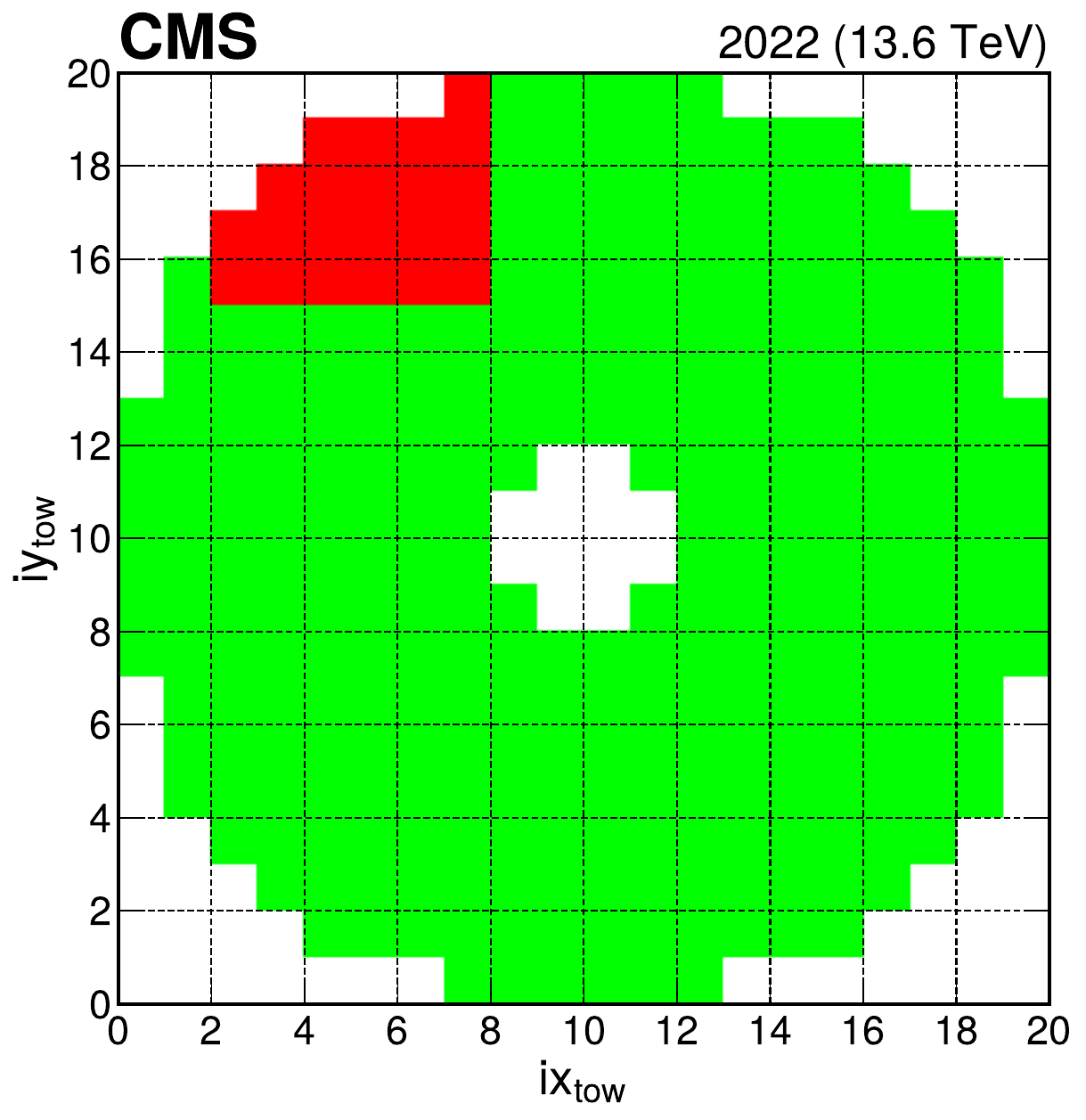}}
}
\caption{Input occupancy images with real anomalies and corresponding AE quality plots for the endcaps.}
\label{fig:realanom_EE}
\end{figure}

Following the tests on fake anomalies, the AE performance is studied on known anomalous data from LHC runs in 2018 and 2022. The input occupancy images with anomalies and the final quality plots from the AE loss maps are illustrated in Figures~\ref{fig:realanom_EB}~and~\ref{fig:realanom_EE}. Figure~\ref{fig:realanom_EB}(a) shows the barrel occupancy map with the supermodule~EB$-$03 in error due to a data unpacking issue from a 2018 run. The AE quality output in Fig.~\ref{fig:realanom_EB}(b) correctly highlights the anomalous supermodule region in red. Figure~\ref{fig:realanom_EB}(c) shows an occupancy map with a region of hot towers and a zero occupancy tower in the center from a 2018 run, and Fig.~\ref{fig:realanom_EB}(d) correctly identifies all the anomalous towers shown in red. It is interesting to note that this error was not detected in the online DQM global quality plots at the time of data taking, while the AE is able to detect it. 

A similar test on real anomalies for the endcaps is illustrated in Fig.~\ref{fig:realanom_EE}. 
An anomaly of more than half of the EE$+$ turned off in a 2018 run (see Fig.~\ref{fig:realanom_EE}(a)) is spotted by the AE quality plot shown in Fig.~\ref{fig:realanom_EE}(b).
Some of the green towers in the red region come from the masked, known problematic towers that the AE has learned during training. Figure~\ref{fig:realanom_EE}(c) shows a region of towers turned off in the upper left quadrant of EE$+$ from a run in 2022, and the AE quality plot in Fig.~\ref{fig:realanom_EE}(d) correctly identifies these towers in red.

The AE is thus able to spot various kinds of anomalies at a tower-level granularity using a single threshold. It does not require any set definitions or rules on the type of anomalies that can be detected, which underlines the importance of unsupervised or semi-supervised ML as a powerful and adaptable anomaly detection tool.

\section{Deployment during LHC Run\,3}
\label{sec:deploy}

\begin{figure}[tbp]
\centering{
\subfloat[]{\includegraphics[width=0.52\textwidth, height=3.5cm]{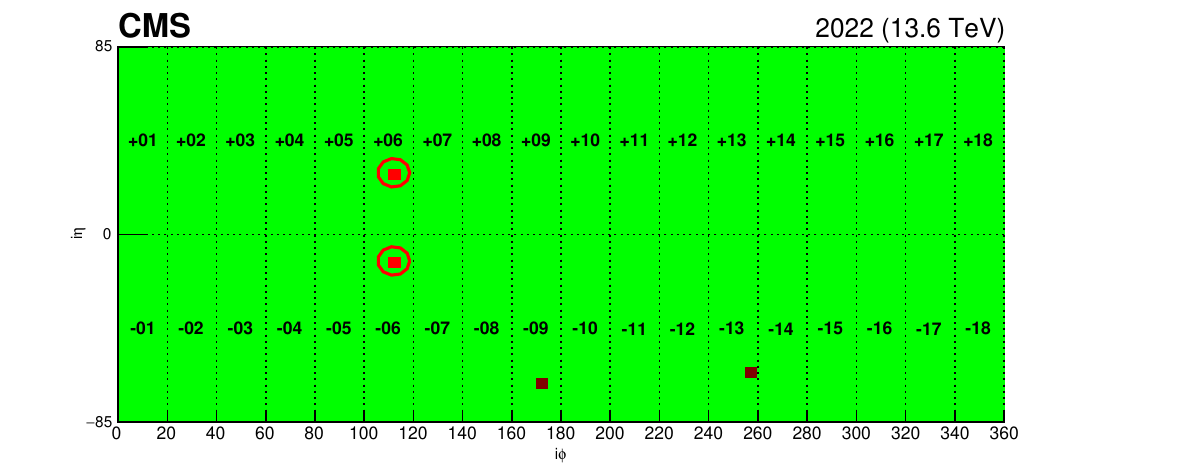}}
\subfloat[]{\includegraphics[width=0.49\textwidth,height=3.5cm]{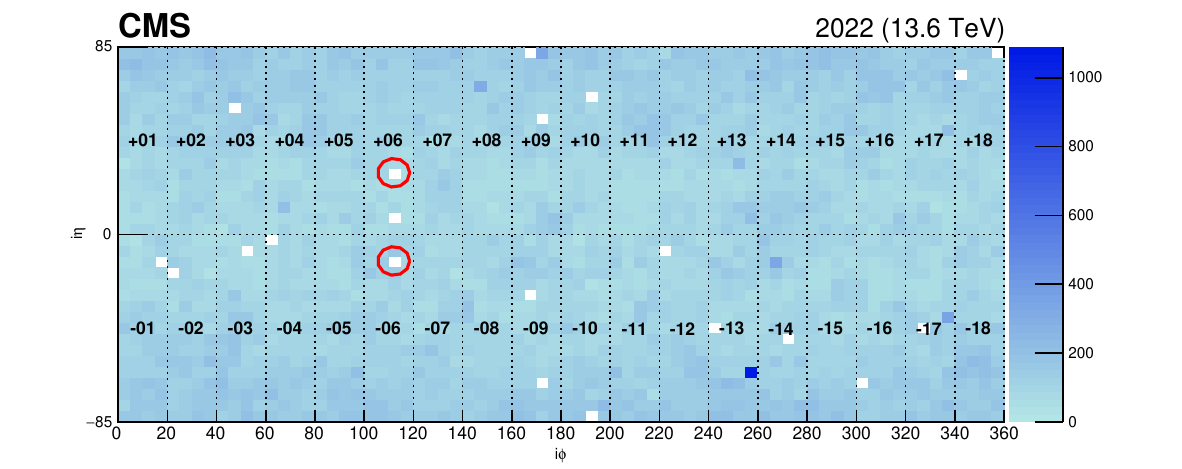}}
}
\caption{From a 2022 Run: (a) ML quality plot in the ECAL DQM from the AE model, with the new bad towers circled. (b) Digi occupancy plot of 1 LS, with the same circled towers with zero occupancy. Four such occupancy plots from four consecutive LSs are summed to make the input to the AE model which results in the quality plot in (a).}
\label{fig:EB-deploy1}
\end{figure}

\begin{figure}[tbp]
\centering{
\subfloat[]{\includegraphics[width=0.4\textwidth]{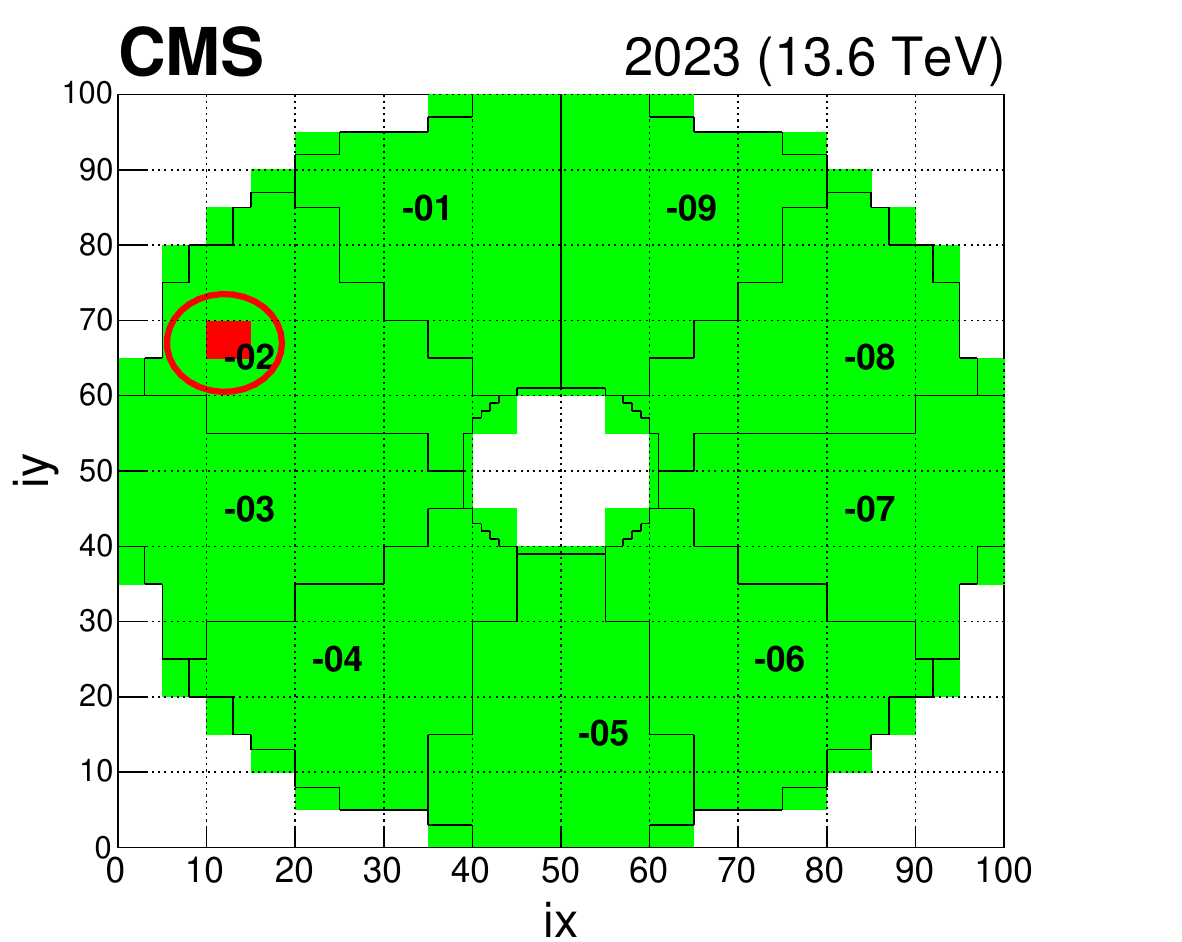}}
\subfloat[]{\includegraphics[width=0.4\textwidth]{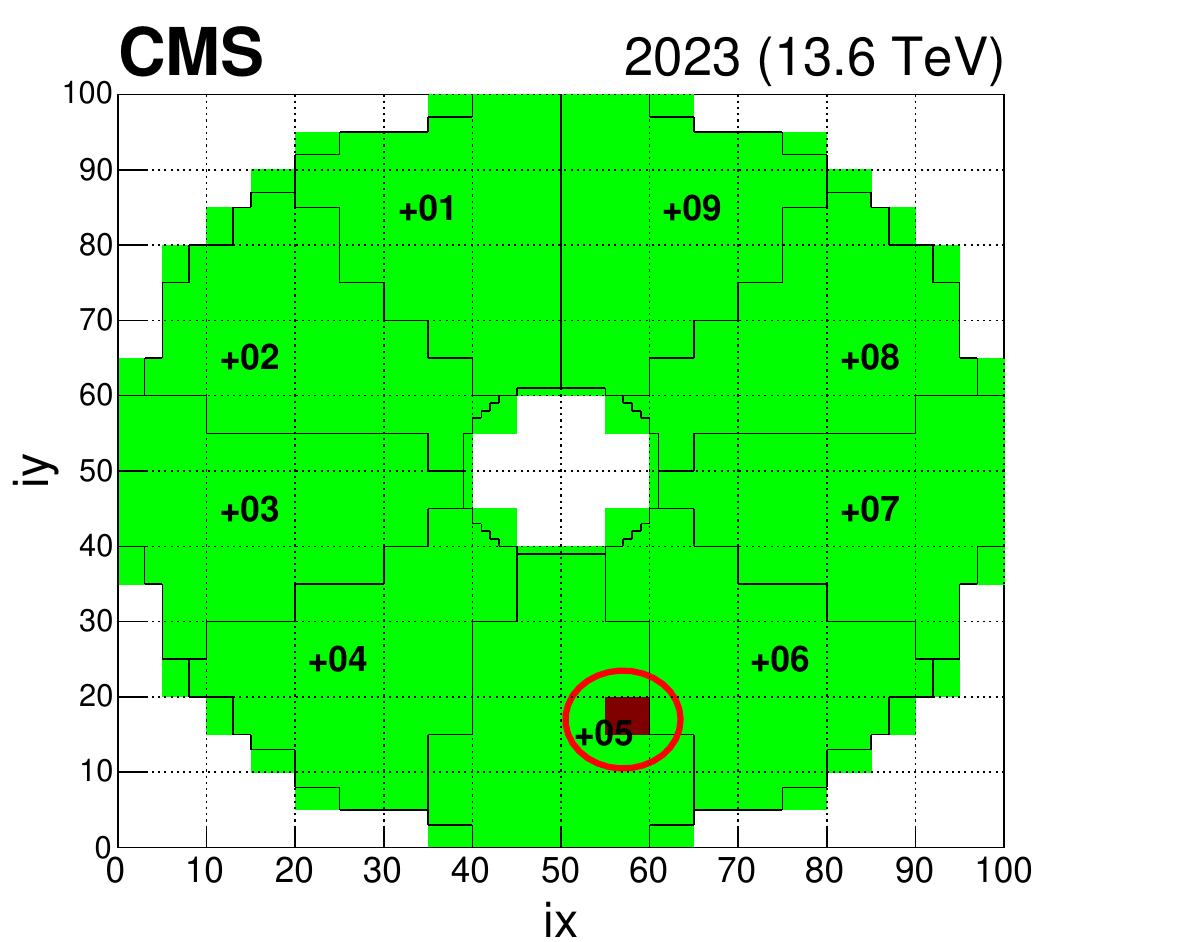}}
\\
\subfloat[]{\includegraphics[width=0.4\textwidth]{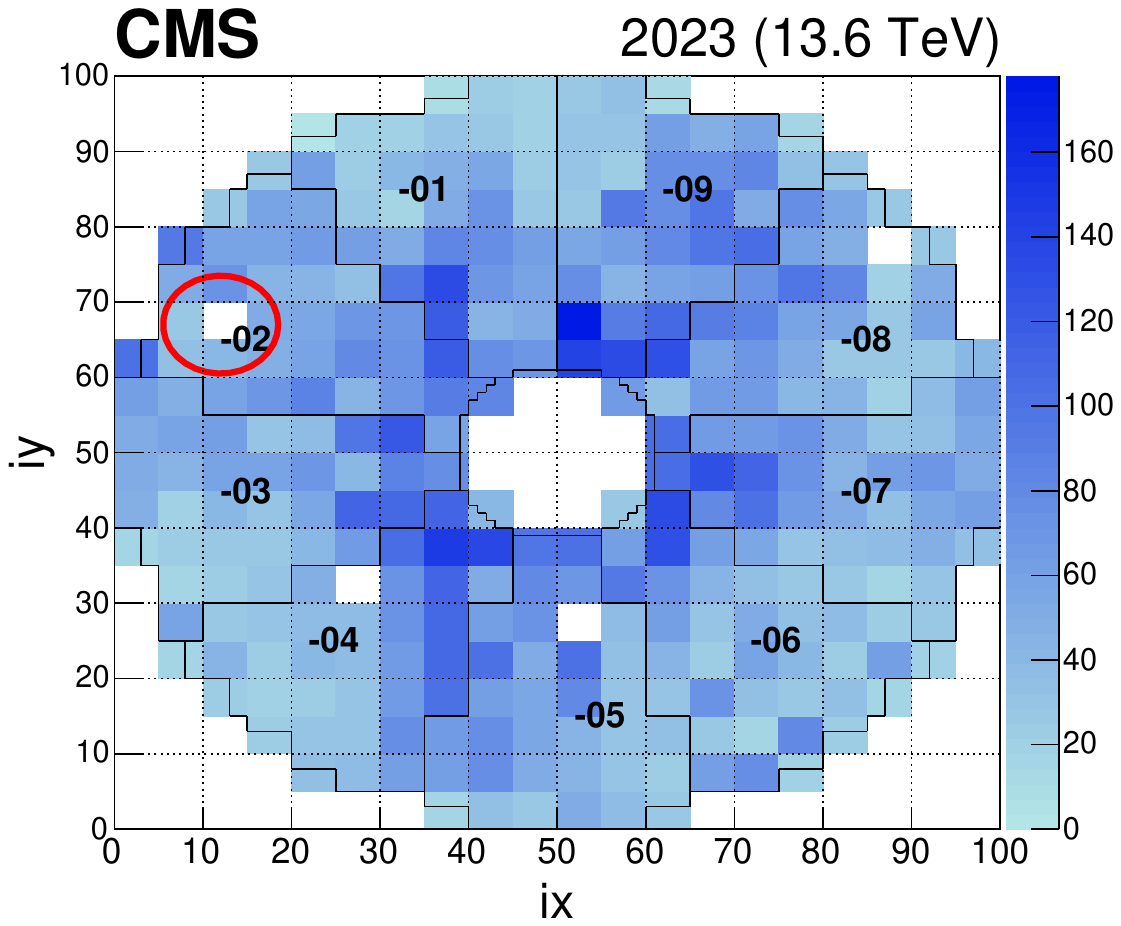}}
\subfloat[]{\includegraphics[width=0.4\textwidth]{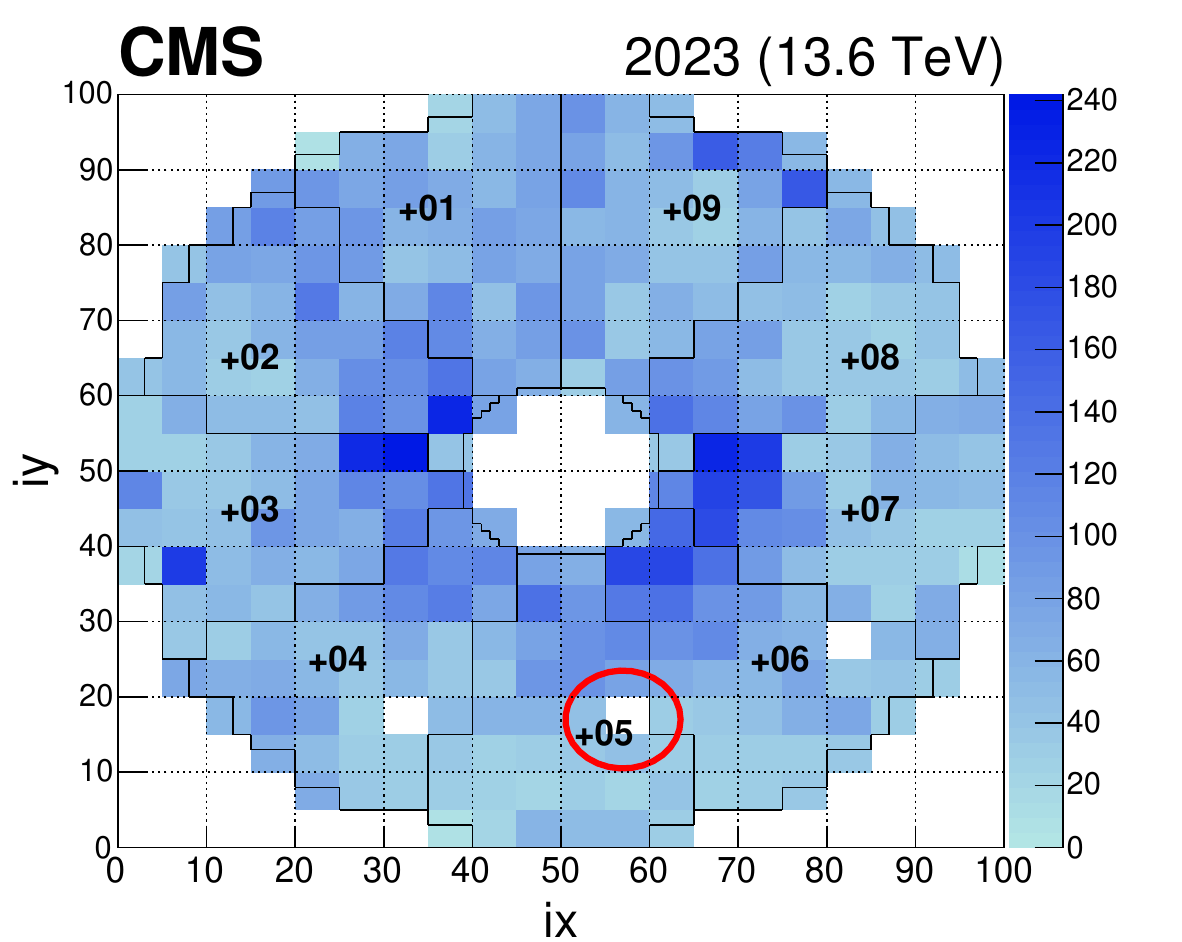}}
}
\caption{From a 2023 Run: ML quality plot in the ECAL DQM from the AE models (a) for the  Endcap EE$-$, with a new bad tower circled and (b) for the Endcap EE$+$ with a known bad tower circled. Digi occupancy plots of 1 LS (c) of EE$-$ and (d) of EE$+$ with the corresponding circled towers with zero occupancy.
Four such occupancy plots from four consecutive LSs are summed to make the input to the AE model which results in the quality plots in (a) and (b).}
\label{fig:EE-deploy1}
\end{figure}

The AE-based anomaly detection system labeled MLDQM has been deployed in the ECAL online DQM workflow in CMSSW for the barrel starting in LHC Run\,3 in 2022 and for the endcaps in 2023. New ML quality plots from the AE (see Fig.~\ref{fig:EB-deploy1}(a) and Fig.~\ref{fig:EE-deploy1}(a)) have been added to the ECAL DQM. The model inference is accomplished using the trained Pytorch models exported to ONNX~\cite{onnx}, which is implemented in the CMS software framework using ONNX Runtime~\cite{onnxruntime}.

The MLDQM models deployed for the barrel and endcaps have shown so far very good performance with Run\,3 data. Figure~\ref{fig:EB-deploy1}(a) illustrates the new quality plot obtained from the inference of the trained AE model for the barrel, using the digi occupancy histogram shown in Fig.~\ref{fig:EB-deploy1}(b) as input to the model.
Similar plots are shown for the endcaps in Figures~\ref{fig:EE-deploy1}(a) and ~\ref{fig:EE-deploy1}(b), which are obtained from the inference of the endcap models with the digi occupancy histograms in Figures~\ref{fig:EE-deploy1}(c) and~\ref{fig:EE-deploy1}(d) as inputs.
The number of events per LS in the digi occupancy map received by the DQM in Run\,3 is approximately 100\,--\,150, smaller than the 500 events per LS used for training. Accordingly, the occupancy maps are summed over four consecutive LSs to collect sufficient statistics. The summed occupancy map is then fed as an input to the AE model, after necessary corrections are applied with respect to the number of events and PU. The resulting loss map then undergoes both the spatial and time corrections. Six consecutive loss maps are used for the time correction during online deployment to minimize false alarms, with the final quality plot essentially accumulated over nine LSs.

From the AE quality plot in Fig.~\ref{fig:EB-deploy1}(a), two circled red towers can be seen in the supermodules~EB+06 and EB$-$06, both corresponding to zero occupancy towers in the input occupancy map as shown in Fig.~\ref{fig:EB-deploy1}(b). The AE quality plot also contains two brown towers that correspond to a zero occupancy tower in~EB$-$09 and a hot tower in~EB$-$13. Both are known problematic towers. 

Similarly, in the AE quality plot for the endcap in Fig.~\ref{fig:EE-deploy1}(a), a red tower can be seen in the sector EE$-$02, corresponding to a zero occupancy tower in the input digi occupancy map in Fig.~\ref{fig:EE-deploy1}(c). From the quality plot shown in Fig.~\ref{fig:EE-deploy1}(b), a brown tower in the sector EE$+$05 corresponds to a zero occupancy tower in the input digi occupancy map in Fig.~\ref{fig:EE-deploy1}(d), which is a known problematic tower.
Other zero occupancy towers in the input occupancy maps that do not show up in the AE quality plots correspond to other known problematic towers that have been present since Run\,2. These known bad towers are learned by the AE during the training process.

\subsection{Detecting Degrading Towers}
\begin{figure}[tbp]
\centering{
\subfloat[]{\includegraphics[width=0.52\textwidth]
{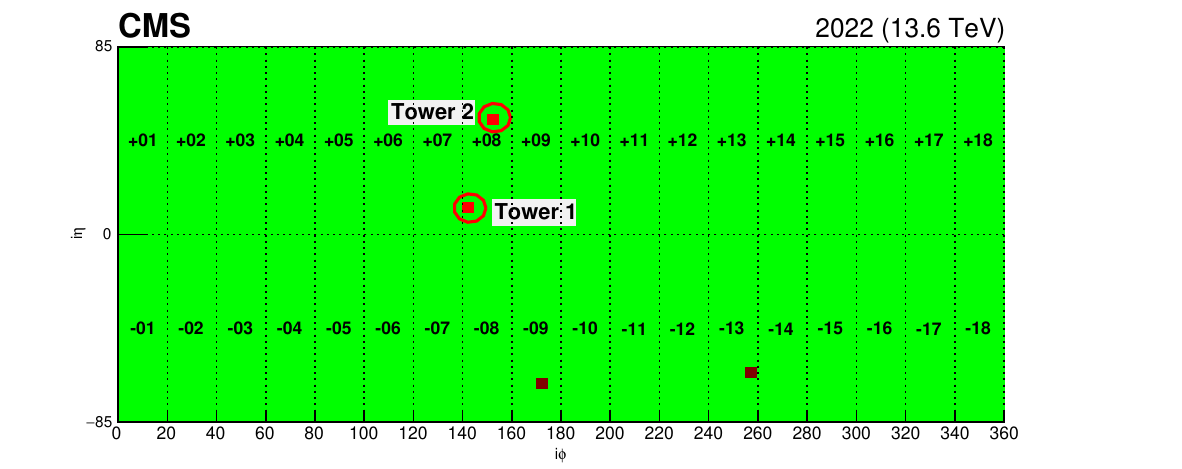}}
\subfloat[]{\includegraphics[width=0.52\textwidth]{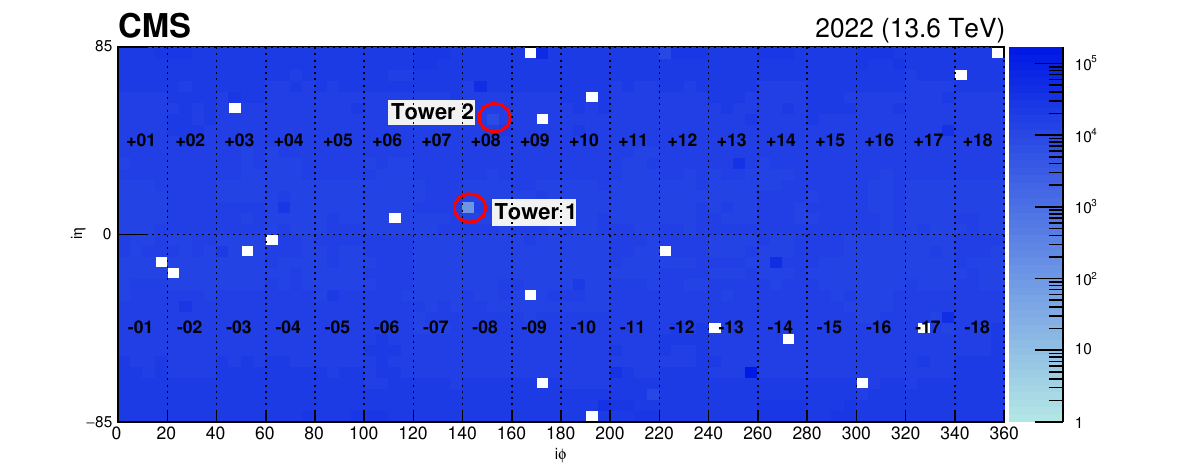}}
\\
\subfloat[]{\includegraphics[width=0.43\textwidth]{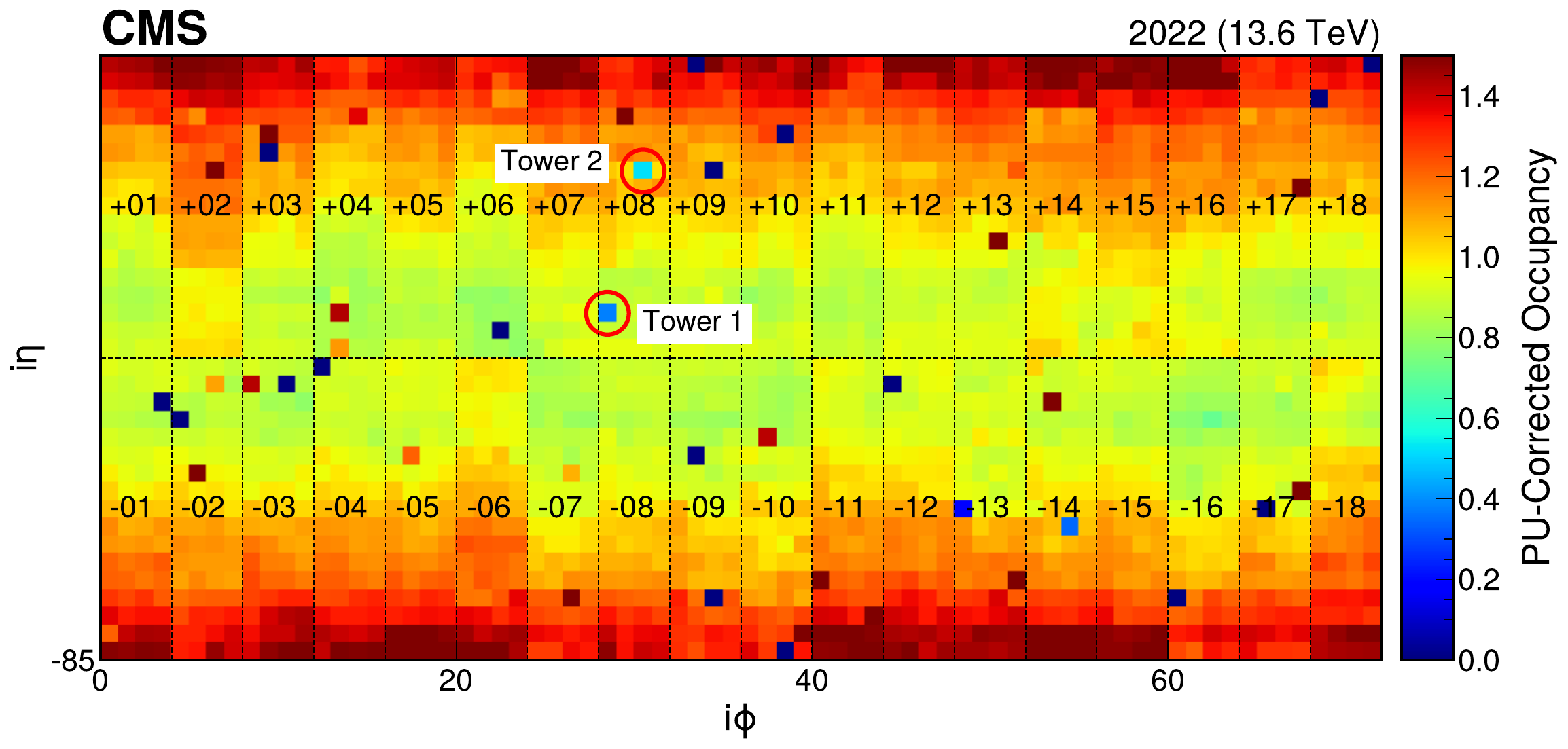}}
}
\caption{From a 2022 Run (a) ML quality plot, accumulated over 9 LSs, in the ECAL DQM from the AE model, (b) The digi occupancy plot accumulated over all the LSs in the full run from the online DQM, and (c) PU-corrected average occupancy map for several runs in 2022.}
\label{fig:EB-deploy-newbad}
\end{figure}

During the MLDQM deployment, it has been observed that the AE can catch new problematic towers with transient anomalous behaviors, which are hard to detect and can be missed by the existing DQM software and plots.
Figure~\ref{fig:EB-deploy-newbad}(a) shows an AE quality plot with two towers in~EB+08 marked in red. Figure~\ref{fig:EB-deploy-newbad}(b) shows the total digi occupancy map accumulated over all LSs from the entire run in the online DQM. Here, Tower\,1 is visible with very low occupancy compared to other towers, indicating that it is a persistent zero occupancy tower. 
On the other hand, the faint visibility of Tower\,2 reflects that it
likely had zero occupancy in several LSs but not in all, possibly corresponding to a transient anomaly.
This feature is also observed in the occupancy map averaged over several runs in Run\,3, e.g. see Fig.~\ref{fig:EB-deploy-newbad}(c). The low occupancy of Tower\,2 in this offline-produced plot implies that the tower indeed had zero occupancy for many LSs in these runs. It is not a random occurrence as in the case of single-event upsets~\cite{Siddireddy:2018gxt} that frequently happen in the detector and are recovered quickly. 
The ability of the AE to identify degrading channels can be of immense use to detector experts when monitoring the health of the ECAL. MLDQM can be used to keep track of how often a particular tower is flagged as bad by the AE, and a threshold can be defined with this frequency to mask the transient tower proactively.

\section{Summary}
\label{sec:summary}

An Autoencoder (AE) based anomaly detection and localization system has been successfully developed, tested, and deployed for the CMS electromagnetic calorimeter~(ECAL) barrel and endcaps using semi-supervised machine learning. 
Occupancy histograms from the ECAL processed as images are used as the input to the network after normalizing the histograms with respect to pileup. Correction strategies are implemented that utilize the variations in the detector response and the time-dependent nature of anomalies. A novel application of the spatial and time corrections yields an order of magnitude improvement in the AE performance. 
Anomaly tagging thresholds chosen at an estimated anomaly tagging rate of 99\% are obtained from validation datasets with fake anomalies introduced. 
In further validations on real anomalies from 2018 and 2022 CMS runs, the AE-based anomaly detection system is able to spot anomalies of various shapes, sizes, and locations in the detector at a tower-level granularity using a single threshold for each part, e.g. the barrel, of the ECAL. 
The deployment of the AE-based system in the Data Quality Monitoring~(DQM) workflow for LHC Run\,3 shows that the system performs well in detecting anomalies, as well as identifying degrading channels that are missed by the existing DQM plots. The AE-based DQM system complements and strengthens the existing DQM by helping detector experts make more accurate decisions and reduce false alarms. The anomaly detection system using machine learning described in this paper can be generalized and adapted not only to other subsystems of the CMS detector but also to other particle physics experiments for anomaly detection and data quality monitoring.

\begin{acknowledgments}
  We congratulate our colleagues in the CERN accelerator departments for
  the excellent performance of the LHC and thank the technical and
  administrative staffs at CERN and at other CMS institutes for their
  contributions to the success of the CMS effort. In addition, we
  gratefully acknowledge the computing centres and personnel of the
  Worldwide LHC Computing Grid and other centres for delivering so
  effectively the computing infrastructure essential to our
  analyses. Finally, we acknowledge the enduring support for the
  construction and operation of the LHC, the CMS detector, and the
  supporting computing infrastructure provided by the following funding
  agencies: SC (Armenia), BMBWF and FWF (Austria); FNRS and FWO
  (Belgium); CNPq, CAPES, FAPERJ, FAPERGS, and FAPESP (Brazil); MES and
  BNSF (Bulgaria); CERN; CAS, MoST, and NSFC (China); MINCIENCIAS
  (Colombia); MSES and CSF (Croatia); RIF (Cyprus); SENESCYT (Ecuador);
  MoER, ERC PUT and ERDF (Estonia); Academy of Finland, MEC, and HIP
  (Finland); CEA and CNRS/IN2P3 (France); BMBF, DFG, and HGF (Germany);
  GSRI (Greece); NKFIH (Hungary); DAE and DST (India); IPM (Iran); SFI
  (Ireland); INFN (Italy); MSIP and NRF (Republic of Korea); MES
  (Latvia); LAS (Lithuania); MOE and UM (Malaysia); BUAP, CINVESTAV,
  CONACYT, LNS, SEP, and UASLP-FAI (Mexico); MOS (Montenegro); MBIE (New
  Zealand); PAEC (Pakistan); MES and NSC (Poland); FCT (Portugal); JINR
  (Dubna); MON, RosAtom, RAS, RFBR, and NRC KI (Russia); MESTD (Serbia);
  MCIN/AEI and PCTI (Spain); MOSTR (Sri Lanka); Swiss Funding Agencies
  (Switzerland); MST (Taipei); MHESI and NSTDA (Thailand); TUBITAK and
  TENMAK (Turkey); NASU (Ukraine); STFC (United Kingdom); DOE and NSF
  (USA). The principal authors of this study are grateful for the work
  of Amy Germer on an initial version of the autoencoder-based system
  for the ECAL endcap sections of the CMS detector.
\end{acknowledgments}

\bibliography{ECAL-ML4DQM}
\cleardoublepage \appendix\section{The CMS ECAL Collaboration \label{app:collab}}\begin{sloppypar}\hyphenpenalty=5000\widowpenalty=500\clubpenalty=5000{D.~Abadjiev}$^{1}$,
{T.~Adams\cmsorcid{0000-0001-8049-5143}}$^{2}$,
{P.~Adzic\cmsorcid{0000-0002-5862-7397}}$^{3,aa}$,
{M.~Ahmad\cmsorcid{0000-0001-9933-995X}}$^{4}$,
{C.~Amendola\cmsorcid{0000-0002-4359-836X}}$^{5}$,
{M.B.~Andrews\cmsorcid{0000-0001-5537-4518}}$^{6}$,
{R.~Arcidiacono\cmsorcid{0000-0001-5904-142X}}$^{7a,7c}$,
{S.~Argiro\cmsorcid{0000-0003-2150-3750}}$^{7a,7b}$,
{A.~Askew\cmsorcid{0000-0002-7172-1396}}$^{2}$,
{E.~Auffray\cmsorcid{0000-0001-8540-1097}}$^{8}$
% ...

\vskip\cmsinstskip
$^{1}$Northeastern University, Boston, Massachusetts, USA\\
$^{2}$Florida State University, Tallahassee, Florida, USA\\
$^{3}$VINCA Institute of Nuclear Sciences, University of Belgrade, Belgrade, Serbia\\
$^{4}$Texas A\&M University, College Station, Texas, USA\\
$^{5}$IRFU, CEA, Universit\'{e} Paris-Saclay, Gif-sur-Yvette, France\\
$^{6}$Carnegie Mellon University, Pittsburgh, Pennsylvania, USA\\
$^{7a}$INFN Sezione di Torino, Torino, Italy\\
$^{7b}$Universit\`{a} di Torino, Torino, Italy\\
$^{7c}$Universit\`{a} del Piemonte Orientale, Novara, Italy\\
$^{8}$CERN, European Organization for Nuclear Research, Geneva, Switzerland\\
\begin{flushleft}
% ...
$^{aa}$Also at Faculty of Physics, University of Belgrade, Belgrade, Serbia\\
% ...
\end{flushleft}

\end{sloppypar}
\end{document}